\documentclass[12pt]{article}
\usepackage{}
\usepackage{amssymb}
\usepackage{amsmath}
\usepackage{graphicx}
\usepackage{indentfirst}
\usepackage{cite}

\allowdisplaybreaks

\begin{document}

\title{Temperature-dependent cross sections \\
for charmonium dissociation in collisions \\
with kaons and $\eta$ mesons in hadronic matter}
\author{Shi-Tao Ji, Zhen-Yu Shen, Xiao-Ming Xu}
\date{}
\maketitle \vspace{-1cm}
\centerline{Department of Physics, Shanghai University, Baoshan, 
Shanghai 200444, China}

\begin{abstract}
We study kaon-charmonium and $\eta$-charmonium dissociation reactions. 
The $K$-charmonium dissociation and the $\eta$-charmonium dissociation
include the following 27 reactions:
$K J/\psi\to\bar{D}^* D^+_s$, $\bar{D} D^{*+}_s$ and $\bar{D}^* D^{*+}_s$;
$K \psi'\to\bar{D}^* D^+_s$, $\bar{D} D^{*+}_s$ and $\bar{D}^* D^{*+}_s$;
$K \chi_c \to\bar{D}^* D^+_s$, $\bar{D} D^{*+}_s$ and $\bar{D}^* D^{*+}_s$;
$\eta J/\psi\to\bar{D}^{*} D$, $\bar{D} D^{*}$, $\bar{D}^{*} D^{*}$,
$D^{*-}_{s} D_s^+$, $D_s^- D^{*+}_s$ and $D^{*-}_{s} D^{*+}_{s}$;
$\eta \psi'\to\bar{D}^{*} D$, $\bar{D} D^{*}$, $\bar{D}^{*} D^{*}$,
$D^{*-}_s D_s^+$, $D_s^- D^{*+}_s$ and $D^{*-}_{s} D^{*+}_{s}$;
$\eta \chi_{c}\to\bar{D}^{*} D$, $\bar{D} D^{*}$, $\bar{D}^{*} D^{*}$,
$D^{*-}_s D_s^+$, $D_s^- D^{*+}_s$ and $D^{*-}_{s} D^{*+}_{s}$.
Cross sections for the reactions 
are calculated in the Born approximation, in the 
quark-interchange mechanism and with a temperature-dependent quark potential.
The temperature dependence of peak cross sections of endothermic reactions is
linked to the temperature dependence of quark-antiquark relative-motion wave 
functions, meson masses and the quark potential. 
Although the $\eta$ meson and kaon
have similar masses, the energy and temperature dependence of the
$\eta$-charmonium dissociation cross sections are quite different from those of
the $K$-charmonium dissociation cross sections. Using the $\eta$-charmonium
and $\pi$-charmonium dissociation cross sections,
we calculate the ratio of the corresponding dissociation rates in hadronic
matter and we find that such rates are comparable at low $J/\psi$ momenta.
\end{abstract}



\section{Introduction}
\label{Introduction}

Nucleon-$J/\psi$ dissociation cross sections 
have been calculated in Refs. \cite{KS,Peskin}
from the gluon-$J/\psi$ dissociation cross section obtained in
short-distance QCD. $J/\psi$ can be 
dissociated by a gluon with energy larger than the quark-antiquark 
binding energy of $J/\psi$ \cite{KS,Peskin,XKSW}. As the nucleon energy 
increases, it generates an increasing number of gluons overcoming the binding
so that $J/\psi$ can be 
dissociated more easily. Therefore, the nucleon-$J/\psi$ dissociation cross 
section increases with total energy of $N$ and $J/\psi$ in the 
centre-of-mass frame. This is a result of the short-distance
approach in which the operator product expansion of perturbative QCD is applied
to heavy quarkonia of small sizes \cite{KS,Peskin}. 
A similar energy dependence of $N+J/\psi$
and $\pi +J/\psi$ dissociation cross sections has been obtained in Ref. 
\cite{AGGA} despite hadron mass corrections to the cross-section formula of
Peskin and Bhanot.

Including the Coulomb potential, the Fermi contact term acting as the 
spin-spin 
interaction and a colour-independent confining force active only between 
a quark and an antiquark, cross sections for 
$\pi + J/\psi \to \bar{D}^{*} +D$, $\pi + J/\psi \to \bar{D} +D^*$ and 
$\pi + J/\psi \to \bar{D}^{*}+D^{*}$ have been calculated in Ref. \cite{MBQ}
in the quark-interchange mechanism \cite{BDPK,BS}. The colour 
interaction that includes the colour Coulomb, spin-spin hyperfine and linear 
confining interactions has been considered in Refs. \cite{WSB,BSWX} for the
dissociation of charmonia in collisions with $\pi$ and $\rho$ mesons. 
These quark-model calculations of the quark-interchange mechanism give the 
result that the cross section increases from a threshold energy for every 
endothermic reaction from zero, reaching a maximum and decreasing with total 
energy of the meson and charmonium in the centre-of-mass frame. 
This is the energy dependence of the cross section 
obtained in the quark-interchange approach of the quark potential models 
\cite{MBQ,BDPK,BS,WSB,BSWX}, where the quark interchange mechanism between
the two initial mesons breaks the charmonium. 

A study of 
$\pi + J/\psi \to \bar{D}^* + D$, $\pi + J/\psi \to \bar{D} + D^*$ and 
$\rho + J/\psi \to \bar{D} + D$ at low 
energies of the mesons was initiated in Ref. \cite{MatinyanM}.
Meson exchange between the two initial mesons breaks the charmonium and
Lagrangians with meson couplings are constructed to describe the motion of
meson fields. This is the essence of the meson-exchange approach
\cite{MatinyanM,LK,Haglin,OSL,NNR,MPPR,BG}.
Effective meson Lagrangians with 
different symmetries lead to different cross sections for the same reaction
\cite{LK,Haglin,OSL,NNR,MPPR,BG}. Near threshold energies cross sections
increase for endothermic reactions and decrease for exothermic reactions.
When the total energy of the two initial mesons in the centre-of-mass frame 
rises far away from threshold, the increase or decrease of the cross
sections is due to form factors inserted in three-meson and 
four-meson vertices of Feynman diagrams.

The energy dependence of cross sections for some
hadron-charmonium dissociation reactions in 
vacuum has been studied in the short-distance, the meson-exchange
and the quark-interchange approaches. In hadronic matter not only the 
energy dependence is interesting, but also the temperature dependence of 
hadron-charmonium dissociation cross sections is important \cite{Wong}.
From vacuum to 
hadronic matter the quark potential, mesonic quark-antiquark relative-motion 
wave functions, meson masses and cross sections for $\pi$-charmonium and 
$\rho$-charmonium dissociation change significantly. The temperature dependence
of the cross sections has been shown to be unexpected \cite{ZX}. 
For instance, even though a $\rho$-charmonium reaction is exothermic in vacuum,
it may be endothermic in hadronic matter; the peak cross section of any
$\pi$-charmonium dissociation reaction first
decreases with increasing temperature, but then
increases rapidly when the temperature approaches the critical temperature.
Therefore, we continue here the study of temperature dependence. We choose 
kaon-charmonium dissociation reactions as the first objective of the present
work.
The energy dependence of the cross sections for the reactions in vacuum has 
been studied in Ref. \cite{WSB}, but in hadronic matter it is not available.
The temperature dependence of the kaon-charmonium 
dissociation cross sections has not been studied either.

It has been shown by experiments that the ratios $K^+/\pi^+$, $K^-/\pi^-$ and 
$\eta /\pi^0$ at midrapidity increase with increasing transverse momentum $p_T$
and the increase is visible only in a certain range of transverse momentum.
For central Au+Au collisions at $\sqrt {s_{NN}}=200$ GeV the
ratio $K^+/\pi^+$ ($K^-/\pi^-$) measured by the PHENIX Collaboration is 
availabe from $p_T=0.45$ GeV/$c$ to 1.95 GeV/$c$ and it is 0.234 (0.221)
at $p_T=0.65$ GeV/$c$ and 0.55 (0.55) at $p_T=1.65$ GeV/$c$ at midrapidity 
\cite{PHENIX}. 
For central Pb+Pb collisions at $\sqrt{s_{NN}}=2.76$ TeV the ratio
$(K^++K^-)/(\pi^++\pi^-)$ measured by the ALICE Collaboration increases with
increasing transverse momentum from $p_T=0$, forms a peak at
$p_T \approx 3$ GeV/$c$ and goes to the value 0.45 at $p_T>4$ GeV/$c$
\cite{ALICE}.
At present, for central Au+Au collisions at $\sqrt {s_{NN}}=200$ GeV
the smallest $\eta$ transverse
momentum measured by the PHENIX Collaboration is
1.65 GeV/$c$ and the ratio $\eta /\pi^0$ is 0.46 
at $p_T=1.65$ GeV/$c$ at midrapidity \cite{Sahlmuller}. From the
measured ratios we conclude that the $\eta$ species is as rich as the $K^+$
and $K^-$ species in hadronic matter and we need to study $\eta$-charmonium
dissociation reactions which form the second objective of the present work.
The short-distance approach and the quark-interchange approach have not been
studied for the $\eta$-charmonium dissociation reactions. The $\eta + J/\psi$
dissociation in vacuum was considered in Ref. \cite{Haglin}, but no cross 
sections were presented. However, we will show that 
$\eta + J/\psi$ dissociation reactions need to be considered in the study of
$J/\psi$ suppression in hadronic matter.

This paper is organized as follows. In the next section we present 
cross-section formulas for the charmonium dissociation in collision with a
meson. In Section 3 kaon-charmonium and $\eta$-charmonium dissociation
reactions are listed, numerical cross sections are shown, relevant
discussions are given and the numerical results are parametrized in Appendix A.
Using the $\eta$-charmonium dissociation cross sections and the 
$\pi$-charmonium dissociation cross sections given in Ref. \cite{ZX}, we 
calculate the ratio of the corresponding dissociation rates in hadronic matter.
A summary is in the last section.

\section{Cross-section formulas}
\label{formula}

In quark degrees of freedom the meson-charmonium dissociation is expressed
as $q\bar{q}+c\bar{c}\rightarrow q\bar{c}+c\bar{q}$ where $q$ stands for the
up quark, down quark or strange quark. The flavour of the quark $q$ may be
different from the flavour of the antiquark $\bar q$.
The cross section for $q\bar{q}+c\bar{c}\rightarrow q\bar{c}+c\bar{q}$ is 
\cite{LX}
\begin{equation}
\sigma(S,m_S,\sqrt {s},T) =\frac{1}{32\pi s}\frac{|\vec{P}^{\prime }(\sqrt{s})|
}{|\vec{P}(\sqrt{s})|}\int_{0}^{\pi }d\theta
|\mathcal{M}_{\rm fi} (s,t)|^{2}\sin \theta,\label{eq:sigma}
\end{equation}
where $S$ is the total spin of either the two initial mesons or the two final
mesons, and it is conserved in the reaction;
$m_S$ denotes the magnetic projection quantum number of the total
spin. The Mandelstam variables are
$s=(E_{q\bar{q}}+E_{c\bar{c}})^2-(\vec{P}_{q\bar{q}}+\vec{P}_{c\bar{c}})^2$
and 
$t=(E_{q\bar{q}}-E_{q\bar{c}})^2-(\vec{P}_{q\bar{q}}-\vec{P}_{q\bar{c}})^2$, 
where $P_i=(E_{i},\vec{P}_i)$ is the 
four-momentum of meson $i~(i=q\bar{q},c\bar{c},q\bar{c},c\bar{q})$;
$\theta$ is the angle between the three-dimensional $q\bar q$
momentum $\vec{P}$ and the three-dimensional $q\bar c$ momentum $\vec{P}'$
in the centre-of-mass frame.

$\mathcal{M}_{\rm fi}$ in Eq. (1) stands for the transition amplitude, which 
takes different values for the two forms of meson-meson scattering, 
namely, the prior form in Fig. 1 and the post form in Fig. 2 
\cite{MottM,BBS,WC}. The scattering
in the prior form means that gluon exchange occurs before
quark interchange. The corresponding transition amplitude is
\begin{equation}
{\cal M}_{\rm fi}^{\rm prior} = 
4\sqrt {E_{q\bar{q}} E_{c\bar{c}}E_{q\bar{c}}E_{c\bar{q}}}
\langle\psi_{q\bar {c}}|\langle\psi_{c\bar{q}}|
(V_{q\bar {c}} +V_ {c \bar {q}} +V_ {qc}+V_{\bar {q}\bar {c}})
|\psi_{q\bar {q}}\rangle|\psi_{c\bar {c}}\rangle, \label{eq:taprior}
\end{equation}
where the meson wave functions,
$\psi_{q\bar {c}}$, $\psi_{c\bar{q}}$, $\psi_{q\bar {q}}$ and 
$\psi_{c\bar {c}}$, represent
the products of quark-antiquark relative-motion wave 
functions in momentum space and wave functions in the colour space, spin space
and flavour space. $V_{q\bar c}$ is the potential of $q$ and $\bar c$ and the
meaning of $V_{c\bar q}$ and so on can be understood in the same way.
The scattering in the post form contains gluon exchange after 
quark interchange. The corresponding transition amplitude is
\begin{equation}
{\cal M}_{\rm fi}^{\rm post}  = 
4\sqrt {E_{q\bar{q}} E_{c\bar{c}}E_{q\bar{c}}E_{c\bar{q}}}
\langle\psi_{q\bar {c}}|\langle\psi_{c\bar{q}}|
(V_{q\bar {q}} +V_ {c \bar {c}} +V_ {qc}+V_{\bar {q}\bar {c}})
|\psi_{q\bar {q}}\rangle|\psi_{c\bar {c}}\rangle. \label{eq:tapost}
\end{equation}

The potential in momentum space in Eqs. (2) and (3) is the Fourier transform
of the potential provided in Ref. \cite{ZX},
\begin{equation}
V_{ab}(\vec {r}) = V_{\rm si}(\vec {r}) + V_{\rm ss}(\vec {r}),
\end{equation}
where $V_{\rm si} (\vec {r})$ is a central spin-independent 
potential and $V_{\rm ss} (\vec {r})$ is a spin-spin
interaction. The temperature dependence is explicitly expressed in
$V_{\rm si} (\vec {r})$,
\begin{equation}
V_{\rm {si}}(\vec {r}) =
-\frac {\vec {\lambda}_a}{2} \cdot \frac {\vec{\lambda}_b}{2}
\frac{3}{4} D \left[ 1.3- \left( \frac {T}{T_{\rm c}} \right)^4 \right]
\tanh (Ar)
+ \frac {\vec {\lambda}_a}{2} \cdot \frac {\vec {\lambda}_b}{2}
\frac {6\pi}{25} \frac {v(\lambda r)}{r} \exp (-Er),
\label{eq:T vsi}
\end{equation}
where $D=0. 7$ GeV, $T_{\rm c}=0.175$ GeV, 
$A=1.5[0.75+0.25 (T/{T_{\rm c}})^{10}]^6$ GeV, $E=0. 6$ GeV and
$\lambda=\sqrt{3b_0/16\pi^2 \alpha'}$ in which $\alpha'=1.04$ GeV$^{-2}$
and $b_{0}=11-\frac{2}{3}N_{f}$ with 
$N_{f}=4$; $\vec {\lambda}_a$ are the Gell-Mann matrices 
for the colour generators of constituent $a$. The dimensionless function
$v(x)$ \cite{BT} is
\begin{eqnarray}
v(x)=\frac
{4b_0}{\pi} \int^\infty_0 \frac {dQ}{Q} \left[\rho (\vec {Q}^2) -\frac {K}{\vec
{Q}^2}\right] \sin \left(\frac {Q}{\lambda}x\right),
\label{eq:v}
\end{eqnarray}
where $K=3/16\pi^2\alpha'$
and $\rho (\vec {Q} ^2)$ is the physical running coupling constant 
at the gluon momentum $\vec {Q}$. 

Lattice QCD calculations gave a 
temperature-dependent quark potential at intermediate and large distances
\cite{KLP}.
In contrast to the linear confinement in vacuum, the quark potential
at large distances is independent of the distance and is a plateau. With
increasing temperature the plateau becomes lower and lower and confinement
gets weaker and weaker. From a large distance to an intermediate distance the
variation of the potential with respect to temperature gets smaller. At
$r \approx 0.3$ fm the variation disappears. The lattice QCD results (the above
potential) \cite{KLP} give the spin-independent
potential in Eq. (5) at intermediate and large 
distances. At short distances quark interaction is described by 
perturbative QCD in vacuum. The spin-independent
potential in Eq. (5) at short distances is thus given by
one-gluon exchange plus perturbative one- and two-loop corrections in
vacuum \cite{BT}. The temperature dependence of the potential 
in Eq. (5) comes from the lattice QCD results.
The potential well fits the lattice QCD results at $T/T_{\rm c}>0.55$
\cite{ZXG} and is constrained by perturbative QCD at short distances and the
lattice QCD results at intermediate and large distances. We may
adjust the parameters $A$ and $E$ to get a good fit of the lattice QCD
results, for example, reduce $A$ by 5\% and increase $E$ by 10\% or increase
$A$ by 5\% and reduce $E$ by 10\%. But the changes of the two parameters lead
to very small changes of meson masses and cross sections
for meson-meson reactions \cite{ZXG}. Hence,
the potential has the uncertainty of the
values of $A$ and $E$, but the uncertainty causes very small changes in meson
mass and in cross section. The expression 
$\frac {\vec {\lambda}_a}{2} \cdot \frac {\vec {\lambda}_b}{2}
\frac {6\pi}{25} \frac {v(\lambda r)}{r}$ in the second term of Eq. (5) is
obtained in perturbative QCD and arises from one-gluon exchange plus
perturbative one- and two-loop corrections in vacuum \cite{BT}. The factor
$\exp (-Er)$ assures that the potential at small distances is given by 
perturbative QCD in vacuum. The second term in Eq. (5) is independent of 
temperature.

From a correlator of a very heavy quark-antiquark pair a time-dependent
potential was obtained in quenched lattice QCD \cite{RHS,BR}. Only at very 
large times and at $T<T_{\rm c}$ the potential agrees with the free energy in 
the Coulomb gauge. Using the free energy from lattice calculations as the
potential of a charm quark and a charm antiquark in the Schr\"odinger equation
correctly describes the nonrelativistic wave function of $J/\psi$ and 
reproduces
the $J/\psi$ mass from the QCD sum rule in the vicinity of the critical 
temperature \cite{LMSK}. When the system's
temperature is smaller than the critical temperature, the product of the
temperature and a meson's entropy is small or negligible in comparison with 
the quark-antiquark free energy and the free energy can be 
taken as the quark-antiquark potential to a good approximation \cite{SX}.
The three methods suggest that the free energy obtained in the lattice QCD
calculations can be taken as the quark potential in hadronic matter.

The spin-spin interaction $V_{\rm ss} (\vec {r})$ arises from
perturbative one-gluon exchange plus one- and two-loop corrections \cite{Xu}
and includes relativistic effects \cite{BS,BSWX,GI}:
\begin{eqnarray}
V_{\rm ss}(\vec {r})=
- \frac {\vec {\lambda}_a}{2} \cdot \frac {\vec {\lambda}_b}{2}
\frac {16\pi^2}{25}\frac{d^3}{\pi^{3/2}}\exp(-d^2r^2) \frac {\vec {s}_a \cdot 
\vec {s} _b} {m_am_b}
+ \frac {\vec {\lambda}_a}{2} \cdot \frac {\vec {\lambda}_b}{2}
  \frac {4\pi}{25} \frac {1} {r}
\frac {d^2v(\lambda r)}{dr^2} \frac {\vec {s}_a \cdot \vec {s}_b}{m_am_b} ,
\label{eq:T vss}
\end{eqnarray}
where $\vec {s}_a$ and $m_a$ are the spin and mass of 
constituent $a$, respectively, and the quantity $d$ is related to quark
masses by
\begin{eqnarray}
d^2=\sigma_{0}^2\left[\frac{1}{2}+\frac{1}{2}\left(\frac{4m_a m_b}{(m_a+m_b)^2}
\right)^4\right]+\sigma_{1}^2\left(\frac{2m_am_b}{m_a+m_b}\right)^2,
\label{eq:d}
\end{eqnarray}
where $\sigma_0=0.15$ GeV and $\sigma_1=0.705$.

Finally, the unpolarised cross section is
\begin{eqnarray}
\sigma ^{\rm unpol} (\sqrt {s}, T) & = &
\frac{1}{(2S_{q\bar q} +1) (2S_ {c\bar c}+1) (2L_ {c\bar c}+1)}
\sum \limits_ {SL_{c\bar{c}z}} (2S+1)
\nonumber\\
&&\times\frac{\sigma^{\rm prior}(S,m_S,\sqrt {s},T)+\sigma^{\rm post}
(S,m_S,\sqrt {s},T)}{2},  \label{eq:sigma unpol}
\end{eqnarray}
where $S_{q\bar q}$ ($S_{c\bar c}$) is the $q\bar q$ ($c\bar c$) spin and
$\sigma^{\rm prior}$ ($\sigma^{\rm post}$) is obtained by the 
replacement of $\mathcal{M}_{\rm fi}$ in Eq. (1) by
$\mathcal{M}_{\rm fi}^{\rm prior}$ ($\mathcal{M}_{\rm fi}^{\rm post}$).
$L_{c\bar{c}z}$ is the magnetic projection quantum number of the orbital
angular momentum $L_{c\bar{c}}$ of the meson $c\bar c$.

\section{Reactions, results and discussions}
\label{result}

Given the charm quark mass $m_c=1.51$ GeV, the up and down quark masses
$m_u=m_d=0.32$ GeV and the strange quark mass $m_s=0.5$ GeV, the Schr\"odinger
equation with the potential in Eq. (4) at $T=0$ reproduces \cite{ZX} the
experimental masses of $J/\psi$, $\psi'$, $\chi_c$, the pion, the rho, the 
kaon, the vector kaon, the $\eta$ meson, the charmed mesons and the charmed 
strange mesons
\cite{PDG}. The potential at $T=0$ and the mesonic quark-antiquark
relative-motion wave functions obtained from the Schr\"odinger equation
reproduce the experimental data of $S$-wave $I=2$ elastic phase shifts for 
$\pi \pi$ scattering in vacuum \cite{Colton,Durusoy,Hoogland,Losty}.

\subsection{Dissociation reactions}
\label{dissociation}

We establish the notation 
$ K= \left( \begin{array}{c} K^+ \\ K^0 \end{array} \right) $,
$\bar{K}= \left( \begin{array}{c} \bar{K}^0 \\ K^- \end{array} \right)$,
$ D= \left( \begin{array}{c} D^+ \\ D^0 \end{array} \right) $,
$\bar{D}= \left( \begin{array}{c} \bar{D}^0 \\ D^- \end{array} \right)$,
$ D^*= \left( \begin{array}{c} D^{*+} \\ D^{*0} \end{array} \right) $ and
$\bar{D}^*= \left( \begin{array}{c} \bar{D}^{*0} \\ D^{*-} \end{array} 
\right)$. The $K$-charmonium dissociation includes the reactions:
\begin{displaymath}
K+J/\psi\to\bar{D}^*+D^+_s,~~~K+J/\psi\to\bar{D}+D^{*+}_s,
~~~K+J/\psi\to\bar{D}^*+D^{*+}_s,
\end{displaymath}
\begin{displaymath}
K+\psi'\to\bar{D}^*+D^+_s,~~~K+\psi'\to\bar{D}+D^{*+}_s,
~~~K+\psi'\to\bar{D}^*+D^{*+}_s,
\end{displaymath}
\begin{displaymath}
K+\chi_c \to\bar{D}^*+D^+_s,~~~K+\chi_c \to\bar{D}+D^{*+}_s,
~~~K+\chi_c \to\bar{D}^*+D^{*+}_s.
\end{displaymath}
The $\bar K$-charmonium dissociation includes
\begin{displaymath}
\bar{K}+J/\psi\to D_s^{*-}+D,~~~\bar{K} + J/\psi\to D_s^- +D^*,
~~~\bar{K}+J/\psi\to D_s^{*-}+D^*,
\end{displaymath}
\begin{displaymath}
\bar{K}+\psi'\to D_s^{*-}+D,~~~\bar{K}+\psi'\to D_s^- +D^*,
~~~\bar{K}+\psi'\to D_s^{*-}+D^*,
\end{displaymath}
\begin{displaymath}
\bar{K}+\chi_c \to D_s^{*-}+D,~~~\bar{K}+\chi_c \to D_s^- +D^*,
~~~\bar{K}+\chi_c \to D_s^{*-}+D^*.
\end{displaymath}
Cross sections for the $\bar {K} - \rm{charmonium}$ dissociation reactions are 
obtained from the $K - \rm{charmonium}$ dissociation reactions, for instance,
the cross section for $\bar{K}+J/\psi\to D_s^- +D^*$ is identical to the cross
section for $K+J/\psi\to\bar{D}^*+D^+_s$. 
Since the $\eta$ meson has the components of $u\bar u$,
$d\bar d$ and $s\bar s$, the $\eta$-charmonium dissociation includes 
$u\bar u$- or $d\bar d$-induced reactions, viz.
\begin{displaymath}
\eta +J/\psi\to\bar{D}^{*}+D,~~~\eta +J/\psi\to\bar{D}+D^{*},
~~~\eta +J/\psi\to\bar{D}^{*}+D^{*},
\end{displaymath}
\begin{displaymath}
\eta +\psi'\to\bar{D}^{*}+D,~~~\eta +\psi'\to\bar{D}+D^{*},
~~~\eta +\psi'\to\bar{D}^{*}+D^{*},
\end{displaymath}
\begin{displaymath}
\eta +\chi_{c}\to\bar{D}^{*}+D,~~~\eta +\chi_{c}\to\bar{D}+D^{*},
~~~\eta +\chi_{c}\to\bar{D}^{*}+D^{*},
\end{displaymath}
and $s\bar s$-induced reactions,
\begin{displaymath}
\eta +J/\psi\to D^{*-}_{s}+D_s^+,~~~\eta +J/\psi\to D_s^- + D^{*+}_s,
~~~\eta +J/\psi \to D^{*-}_{s}+D^{*+}_{s},
\end{displaymath}
\begin{displaymath}
\eta +\psi'\to D^{*-}_s +D_s^+,~~~\eta +\psi'\to D_s^- +D^{*+}_s,
~~~\eta +\psi'\to D^{*-}_{s} +D^{*+}_{s},
\end{displaymath}
\begin{displaymath}
\eta +\chi_{c}\to D^{*-}_s +D_s^+,~~~\eta +\chi_{c}\to D_s^- +D^{*+}_s,
~~~\eta +\chi_{c}\to D^{*-}_{s}+D^{*+}_{s}.
\end{displaymath}
The valence quark-antiquark wave function of the $\eta$ meson is
\begin{equation}
\psi_\eta =\frac {1}{\sqrt 6}\phi_{u\bar u}\phi_{\rm C}\phi_{\rm S}u\bar {u}
+\frac {1}{\sqrt 6}\phi_{d\bar d}\phi_{\rm C}\phi_{\rm S}d\bar {d}
-\frac {2}{\sqrt 6}\phi_{s\bar s}\phi_{\rm C}\phi_{\rm S}s\bar {s},
\end{equation}
where $\phi_{u\bar u}$ ($\phi_{d\bar d}$, $\phi_{s\bar s}$) is the 
relative-motion wave function of the up (down, strange) quark and the up
(down, strange) antiquark in momentum space; $\phi_{\rm C}$ and $\phi_{\rm S}$
are the colour wave function and the spin wave function, respectively. 
$\phi_{u\bar u}$, $\phi_{d\bar d}$ and $\phi_{s\bar s}$ satisfy
\begin{equation}
\int \frac {d^3p_{u\bar u}}{(2\pi)^3}\phi^*_{u\bar u}(\vec {p}_{u\bar u})
\phi_{u\bar u}(\vec {p}_{u\bar u})=
\int \frac {d^3p_{d\bar d}}{(2\pi)^3}\phi^*_{d\bar d}(\vec {p}_{d\bar d})
\phi_{d\bar d}(\vec {p}_{d\bar d})=
\int \frac {d^3p_{s\bar s}}{(2\pi)^3}\phi^*_{s\bar s}(\vec {p}_{s\bar s})
\phi_{s\bar s}(\vec {p}_{s\bar s})=1,
\end{equation}
where $\vec {p}_{u\bar u}$ ($\vec {p}_{d\bar d}$, $\vec {p}_{s\bar s}$) is the
relative momentum of the up (down, strange) quark and the up (down, strange)
antiquark. The 
first (second, third) term is used in the transition amplitude to obtain the
cross sections $\sigma_{\eta+c\bar{c}\rightarrow u\bar{c}+c\bar{u}}$
($\sigma_{\eta+c\bar{c}\rightarrow d\bar{c}+c\bar{d}}$,
$\sigma_{\eta+c\bar{c}\rightarrow s\bar{c}+c\bar{s}}$) for the
$u\bar u$-induced ($d\bar d$-induced, $s\bar s$-induced) reactions.
The cross section for the $\eta$-charmonium dissociation is
\begin{eqnarray}
\sigma_{\eta c\bar {c}}=
\frac{1}{6}\sigma_{\eta+c\bar{c}\rightarrow u\bar{c}+c\bar{u}}
+\frac{1}{6}\sigma_{\eta+c\bar{c}\rightarrow d\bar{c}+c\bar{d}}
+\frac{2}{3}\sigma_{\eta+c\bar{c}\rightarrow s\bar{c}+c\bar{s}}.
\label{eq:sigmatot1}
\end{eqnarray}
The cross section for the production of $\bar {D}^*D$, $\bar {D}D^*$ or
$\bar {D}^*D^*$ equals the first and second terms while the one for the
production of $D_s^{*-}D_s^+$, $D_s^-D_s^{*+}$ or $D_s^{*-}D_s^{*+}$ equals
the third term.

\subsection{Numerical cross sections and discussions}
\label{discussion}

Solving the Schr\"{o}dinger equation with the potential in Eq. (4), we
obtain temperature-dependent meson masses shown in Figs. 1 and 17 in Ref.
\cite{ZX} and quark-antiquark relative-motion wave functions for 
charmonia, charmed mesons, charmed strange mesons and the mesons in the 
ground-state pseudoscalar octet and the ground-state vector nonet. The 
quark-antiquark relative-motion wave functions used in
the transition amplitude are the Fourier transforms of the relative-motion wave
functions obtained here.  Since the up quark has the same mass as the down 
quark, $\phi_{u\bar {u}}$ equals $\phi_{d\bar {d}}$ but
differs from $\phi_{s\bar {s}}$. Then,
the cross sections $\sigma_{\eta+c\bar{c}\rightarrow u\bar{c}+c\bar{u}}$ 
for the $u\bar u$-induced reactions are the same as
$\sigma_{\eta+c\bar{c}\rightarrow d\bar{c}+c\bar{d}}$ for the $d\bar d$-induced
reactions.

Cross sections for exothermic reactions are infinite at threshold energies
and we start calculating the cross sections 
at $\sqrt{s}=m_{q\bar q}+m_{c\bar c}+\Delta\sqrt{s}$ with 
$\Delta\sqrt{s}=10^{-4}$ GeV. Unpolarised cross sections for the $K$-charmonium
dissociation reactions and 
$\eta$-charmonium dissociation reactions are displayed in Figs. 3-23.
The cross sections for exothermic reactions decrease very rapidly from infinity
when $\sqrt s$ increases from the threshold energies. To indicate this
feature of an exothermic reaction, the rapid decrease
should be displayed. However, when we start plotting the cross sections 
at the threshold energies plus $5 \times {10}^{-3}$ GeV, the decreasing part
of the curve for $\eta + \chi_c \to \bar{D}^* + D$ at $T/T_c=0.9$ in Fig. 16
disappears. Likewise when we start plotting the cross sections at the threshold
energies plus $10^{-3}$ GeV, the decreasing part of the curve for 
$\eta + \psi' \to \bar{D}^* + D$ at $T/T_c=0.65$ in Fig. 14 disappears.
Therefore, we choose the threshold energies plus $10^{-4}$ GeV to start 
plotting the cross sections so that the rapid decrease is visible.

In vacuum the cross section for $KJ/\psi \to \bar{D} D^{*+}_s$ obtained from
the effective meson Lagrangian in Ref. \cite{Haglin} increases with increasing
$\sqrt s$ and reaches a magnitude of about 8 mb. In the present work the cross
section for $KJ/\psi \to \bar{D} D^{*+}_s$ shown by the curve at $T=0$ first
increases to a maximum value of about 0.27 mb with increasing $\sqrt s$
and then decreases. Therefore, the magnitude of the
present cross section is much smaller than the magnitude obtained in Ref. 
\cite{Haglin}. The $\eta J/\psi$ dissociation in vacuum was considered in Ref. 
\cite{Haglin}, but no cross sections were shown.
In Ref. \cite{WSB} cross sections for the production of $\bar {D}^* D_s^+$,
$\bar {D} D_s^{*+}$ and $\bar {D}^* D_s^{*+}$ in the
$K$-charmonium dissociation in vacuum were obtained in the quark-interchange
mechanism. The peak cross section of the endothermic reaction
$KJ/\psi \to \bar{D}^*D_s^+ + \bar{D}D_s^{*+} + \bar{D}^*D_s^{*+}$ is about
0.7 mb at the kinetic energy of the order of 0.44 GeV. The cross section for
the exothermic reaction
$K\psi' \to \bar{D}^*D_s^+ + \bar{D}D_s^{*+} + \bar{D}^*D_s^{*+}$ decreases
very rapidly and then increases slowly to form a wide peak that corresponds to
a cross section of about 1 mb. The exothermic reaction
$K\chi_{c1} \to \bar{D}^*D_s^+ + \bar{D}D_s^{*+} + \bar{D}^*D_s^{*+}$ has a
peak cross section of about 3 mb at the kinetic energy of the order of 0.17
GeV. By comparison, we obtain a peak cross section of about 0.6 mb at the
kinetic energy of the order of 0.44 GeV for
$KJ/\psi \to \bar{D}^*D_s^+ + \bar{D}D_s^{*+} + \bar{D}^*D_s^{*+}$, no peak
cross section for
$K\psi' \to \bar{D}^*D_s^+ + \bar{D}D_s^{*+} + \bar{D}^*D_s^{*+}$ and a peak
cross section of about 3.4 mb at the kinetic energy of the order of 0.08 GeV 
for
$K\chi_{c} \to \bar{D}^*D_s^+ + \bar{D}D_s^{*+} + \bar{D}^*D_s^{*+}$. We note
that we use the average value of the $\chi_{c0}$, $\chi_{c1}$ and $\chi_{c2}$
masses as the $\chi_c$ mass and the average value is a little higher than the
$\chi_{c1}$ mass. The present results are comparable to the $K$-charmonium
dissociation cross sections in Ref. \cite{WSB}. The present work and the
quark-interchange model in Ref. \cite{WSB} have the same colour matrix elements
and the same spin matrix elements, but have different quark potentials that
lead to different
spatial matrix elements in the transition amplitude. Hence, the
energy dependence of the $K$-charmonium dissociation cross sections obtained
in the present work is not the same as that displayed in Ref. \cite{WSB}.

In Figs. 3-11 only $K + \psi^\prime \to \bar{D}^* + D^+_s$, 
$K + \psi^\prime \to \bar{D} + D^{*+}_s$,
$K + \psi^\prime \to \bar{D}^* + D^{*+}_s$,
$K + \chi_c \to \bar{D}^* + D^+_s$ and $K + \chi_c \to \bar{D} + D^{*+}_s$
at $T=0$ are exothermic. While temperature goes from $0.65T_{\rm c}$ 
to $0.75T_{\rm c}$, the increases of the kaon and charmonium radii cause the 
increases of the peak cross sections of the $K + J/\psi$ and
$K + \chi_c$ reactions, but the peak cross sections of the $K + \psi'$
reactions decrease. This relates to the node in the $\psi'$ wave function. The
node leads to cancellation between the negative wave function on the left
of the node and the positive wave function on the right of the node in the
integration involved in the transition amplitude. While the cancellation at
$T=0.75T_{\rm c}$ is larger than at $T=0.65T_{\rm c}$, the peak cross sections
of $K + \psi^\prime \to \bar{D}^* + D^+_s$, 
$K + \psi^\prime \to \bar{D} + D^{*+}_s$ and
$K + \psi^\prime \to \bar{D}^* + D^{*+}_s$
decrease from $T=0.65T_{\rm c}$ to
$0.75T_{\rm c}$. While temperature increases from $T=0.75T_{\rm c}$, the
increase of $|\mathcal M_{\rm fi}|^2$ caused by the slow increases of
the two initial-meson radii cannot overcome the reduction by the weakening
confinement and the peak cross sections thus decrease. However,
the rapid increases of the
kaon and $J/\psi$ radii from $T=0.9T_{\rm c}$ to $0.95T_{\rm c}$ cause the
increase of $|\mathcal M_{\rm fi}|^2$ to overcome the reduction by the
weakening confinement, and the peak cross sections of
$K + J/\psi \to \bar{D}^* + D^+_s$, $K + J/\psi \to \bar{D} + D^{*+}_s$ and
$K + J/\psi \to \bar{D}^* + D^{*+}_s$ rise rapidly from $T=0.9T_{\rm c}$
to $0.95T_{\rm c}$.

We need to mention the difference between the cross section for
$K + J/\psi \to \bar{D}^* + D^+_s$ and the one for 
$K + J/\psi \to \bar{D} +D^{*+}_s$. The difference arises from the 
different spins of the charmed mesons and the different
spins of the charmed strange mesons between the two reactions. The masses of 
the $\bar D$, $\bar {D}^*$, $D^+_s$ and $D^{*+}_s$ mesons are different
and the sum of the $\bar{D}^*$ and $D^+_s$ masses
is smaller than the sum of the 
$\bar D$ and $D^{*+}_s$ masses. Hence, the endothermic reaction 
$K + J/\psi \to \bar{D}^* + D^+_s$ takes place more easily than 
$K + J/\psi \to \bar{D} + D^{*+}_s$ and at $T/T_{\rm c}$=0 (0.65, 0.75, 0.85, 
0.9, 0.95) the peak cross section of 
$K + J/\psi \to \bar{D}^* + D^+_s$ is 18\% (47\%, 42\%, 11\%, 135\%, 95\%)
larger than the one of 
$K + J/\psi \to \bar{D} + D^{*+}_s$. The difference between the cross section
for $K + \psi' \to \bar{D}^* + D^+_s$ and the one for
$K + \psi' \to \bar{D} + D^{*+}_s$ 
($K + \chi_{\rm c} \to \bar{D}^* + D^+_s$ and 
$K + \chi_{\rm c} \to \bar{D} + D^{*+}_s$) can be similarly understood.

The cross sections for the $\eta$-charmonium dissociation reactions
have the following behaviour. As shown in Figs. 12-17,
only three reactions $\eta + J/\psi \to \bar{D}^* + D$,
$\eta + J/\psi \to \bar{D} + D^*$  and  $\eta + J/\psi \to \bar{D}^* + D^*$
are endothermic below some temperature. The peak cross sections of the 
reactions increase with temperature from $T=0$ to $0.85T_{\rm c}$. As
shown in Figs. 18-23, only three reactions 
$\eta + \psi^\prime \to D^{*-}_s + D^+_s$, 
$\eta + \psi^\prime \to D^-_s + D^{*+}_s$ and
$\eta + \psi^\prime \to D^{*-}_s + D^{*+}_s$ are exothermic below some
temperature. The peak cross sections of the reactions decrease with 
temperature going from $T=0.65T_{\rm c}$ to $T_{\rm c}$. The peak cross 
sections of $\eta + J/\psi \to D^{*-}_s + D_s^+$,
$\eta + J/\psi \to D^-_s + D_s^{*+}$, 
$\eta + J/\psi \to D^{*-}_s + D^{*+}_s$, 
$\eta + \chi_c \to D^{*-}_s + D_s^+$, $\eta + \chi_c \to D^-_s + D_s^{*+}$ and
$\eta + \chi_c \to D^{*-}_s + D^{*+}_s$ first rise and then fall with
temperature increasing in the region $0 \leq T < T_{\rm c}$.

Since the ratio of the $\eta$ mass to the kaon mass is about 1.1 for
$0 \leq T/T_{\rm c} < 1$, i.e. the two masses are close, it is interesting to
compare the peak cross sections of the
endothermic $\eta$-charmonium dissociation reactions with those of the
endothermic kaon-charmonium dissociation reactions. The peak cross
section of $\eta + J/\psi \to \bar{D}^* + D$ at $T/T_{\rm c}=0$ (0.65, 0.75,
0.85) is 1.64 (1.57, 1.55, 12.11) times the peak cross section of
$K + J/\psi \to \bar{D}^* + D^+_s$ at the same temperature. The peak cross
section of $\eta + J/\psi \to \bar{D}^* + D^*$ at $T/T_{\rm c}=0$ (0.65, 
0.75, 0.85) is 0.56 (1.47, 3.87, 49.4) times the peak cross section 
of $K + J/\psi \to \bar{D}^* + D^{*+}_s$. This means that
for $0.6 \leq T/T_{\rm c} < 1$ the cross sections for 
$\eta + J/\psi \to \bar{D}^* + D$ and $\eta + J/\psi \to \bar{D}^* + D^*$ 
around their cross-section peaks are
larger than the cross sections for $K + J/\psi \to \bar{D}^* + D^+_s$ and
$K + J/\psi \to \bar{D}^* + D^{*+}_s$, respectively. Except for
$\eta + J/\psi \to D^{*-}_s + D^{*+}_s$ and 
$K + J/\psi \to \bar{D}^* + D^{*+}_s$ at $T/T_{\rm c}=0.85$,
the peak cross sections of $\eta +{\rm charmonium} \to D^{*-}_{s}+D_s^+$ and
$\eta + {\rm charmonium} \to D^{*-}_{s}+D^{*+}_{s}$
are smaller than the ones of
$K + {\rm charmonium} \to \bar{D}^* + D^+_s$ and
$K + {\rm charmonium} \to \bar{D}^* + D^{*+}_s$, respectively.

Since the $\eta$-charmonium dissociation and the
$\pi$-charmonium dissociation may
produce the same charmed mesons, it is interesting to compare the peak cross
sections of the endothermic $\eta$-charmonium dissociation reactions with those
of endothermic $\pi$-charmonium dissociation reactions. Cross sections for the
$\pi$-charmonium dissociation reactions are given in Ref. \cite{ZX}.
At any temperature between $T/T_{\rm c}=0.6$ and 1 the peak cross section of 
$\eta+J/\psi \to \bar{D}^*+D$ ($\eta+J/\psi \to \bar{D}^*+D^*$) is more than 
1.8
times the one of $\pi+J/\psi \to \bar{D}^*+D$ ($\pi+J/\psi \to \bar{D}^*+D^*$).
In the $\eta$-charmonium dissociation reactions the $s\bar s$ component of the
$\eta$ meson leads to the production of $D^{*-}_s + D^+_s$, $D^-_s + D^{*+}_s$
or $D^{*-}_s + D^{*+}_s$. These charmed strange mesons are not produced in the
$\pi$-charmonium dissociation reactions. Based on the estimate of the
$\eta$-charmonium and $\pi$-charmonium dissociation cross sections, in the
following we evaluate the difference between the dissociation rates of 
charmonium with $\pi$ and $\eta$ in hadronic matter.

In hadronic matter the $\pi$ and $\eta$ mesons satisfy the Bose-Einstein 
distribution
\begin{equation}
f_{q\bar q}(\vec {k})=\frac {1}{e^{\sqrt{m^2_{q\bar q}+\vec{k}^2}/T}-1},
\end{equation}
where $m_{q\bar q}$ is the $q\bar q$ mass. The $q\bar q$ number density is
\begin{equation}
n_{q\bar q}=g_{q\bar q}\int \frac {d^3k}{(2\pi)^3}f_{q\bar q}(\vec {k}),
\end{equation}
where $g_{q\bar q}$ is the spin-isospin degeneracy factor and equals 3 for the
pion and 1 for the $\eta$ meson.
The thermal-averaged meson-charmonium dissociation cross section is
\begin{equation}
<v_{\rm rel}\sigma^{\rm unpol} (\sqrt {s},T)>
=\frac {g_{q\bar q}\int \frac {d^3k}{(2\pi)^3}v_{\rm rel} 
        \sigma^{\rm unpol} (\sqrt {s},T) f_{q\bar q}(\vec {k})}
       {g_{q\bar q}\int \frac {d^3k}{(2\pi)^3}f_{q\bar q}(\vec {k})},
\end{equation}
where $v_{\rm rel}$ is the relative velocity of the $q\bar q$ meson and 
the charmonium. The dissociation rate of the charmonium in the interaction with
the $q\bar q$ meson in hadronic matter is
\begin{equation}
n_{q\bar q}<v_{\rm rel}\sigma^{\rm unpol} (\sqrt {s},T)>,
\end{equation}
which determines charmonium suppression in mesonic matter
\cite{Xu}. The larger the dissociation rate is, the stronger suppression this
mesonic matter causes. The ratio of the dissociation rates of
charmonium in the interactions with $\eta$ and $\pi$ is
\begin{equation}
R_{\eta / \pi}=\frac 
{n_{\eta}<v_{\rm rel}\sigma^{\rm unpol}_{\eta c\bar {c}} (\sqrt {s},T)>}
{n_{\pi}<v_{\rm rel}\sigma^{\rm unpol}_{\pi c\bar {c}} (\sqrt {s},T)>},
\end{equation}
where $n_{\eta}$ and $n_{\pi}$ are the number densities of $\eta$ and $\pi$
mesons, respectively; $\sigma^{\rm unpol}_{\eta c\bar {c}}$ and 
$\sigma^{\rm unpol}_{\pi c\bar {c}}$ are the unpolarised cross sections for a
$\eta$-charmonium dissociation reaction and a $\pi$-charmonium dissociation 
reaction, respectively. We calculate the ratio for the six sets of reactions:
\begin{displaymath}
(1)~~~~~\eta + J/\psi \to \bar {D}^* + D, ~~~~~\pi + J/\psi \to \bar {D}^* + D;
\end{displaymath}
\begin{displaymath}
(2)~~~~~\eta + J/\psi \to \bar {D}^* + D^*, 
~~~~~\pi + J/\psi \to \bar {D}^* + D^*;
\end{displaymath}
\begin{displaymath}
(3)~~~~~\eta + \psi' \to \bar {D}^* + D, ~~~~~\pi + \psi' \to \bar {D}^* + D;
\end{displaymath}
\begin{displaymath}
(4)~~~~~\eta + \psi' \to \bar {D}^* + D^*, 
~~~~~\pi + \psi' \to \bar {D}^* + D^*;
\end{displaymath}
\begin{displaymath}
(5)~~~~~\eta + \chi_c \to \bar {D}^* + D, ~~~~~\pi + \chi_c \to \bar {D}^* + D;
\end{displaymath}
\begin{displaymath}
(6)~~~~~\eta + \chi_c \to \bar {D}^* + D^*, 
~~~~~\pi + \chi_c \to \bar {D}^* + D^*.
\end{displaymath}
Results are shown in Figs. 24-29. The two reactions in each set have the same
final states. The ratio decreases with charmonium momentum increase. For
$\eta + \psi' \to \bar {D}^* + D$ and $\pi + \psi' \to \bar {D}^* + D$ the 
ratio in Fig. 26 is smaller than 0.018. For $\eta + \psi' \to \bar {D}^* + D^*$
and $\pi + \psi' \to \bar {D}^* + D^*$ the ratio in Fig. 27 is smaller than 
0.82. For $\eta + \chi_c \to \bar {D}^* + D$ and 
$\pi + \chi_c \to \bar {D}^* + D$ the ratio in Fig. 28 is smaller than 0.08. 
For $\eta + \chi_c \to \bar {D}^* + D^*$ and 
$\pi + \chi_c \to \bar {D}^* + D^*$ the ratio in Fig. 29 is smaller than 0.72.
For $\eta + J/\psi \to \bar {D}^* + D$ and 
$\pi + J/\psi \to \bar {D}^* + D$ the ratio at $T/T_{\rm c}$=0.65, 0.75, 0.9
and 0.95 in Fig. 24 is smaller than 0.73 and the ratio at $T/T_{\rm c}$=0.85 is
larger than 1 when the $J/\psi$ momentum is smaller than 0.7 GeV/$c$.
For $\eta + J/\psi \to \bar {D}^* + D^*$ and 
$\pi + J/\psi \to \bar {D}^* + D^*$ the ratio at $T/T_{\rm c}$=0.65 and 0.95
in Fig. 25
is smaller than 0.51 and the ratio at $T/T_{\rm c}$=0.75, 0.85 and 0.9 is
larger than 1 when the $J/\psi$ momentum is smaller than 0.6 GeV/$c$,
3.3 GeV/$c$ and 6.6 GeV/$c$, respectively.
Therefore, compared to the dissociation rate of $\psi'$ ($\chi_c$) with
$\pi$, the dissociation rate of $\psi'$ ($\chi_c$) with
$\eta$ may be neglected; but the dissociation rate of $J/\psi$ with
$\eta$ is comparable to the dissociation rate of $J/\psi$ with
$\pi$ at low $J/\psi$ momenta. To study the $J/\psi$ suppression in
hadronic matter, the $\eta + J/\psi$ dissociation reactions need to be
considered.

\section{Summary}
\label{Summary}

We have studied the production of $\bar{D}^*D^+_s$, $\bar{D}D^{*+}_s$ and
$\bar{D}^*D^{*+}_s$ in the $K$-charmonium dissociation and the production of
$\bar{D}^*D$, $\bar{D}D^*$, $\bar{D}^*D^*$, $D^{*-}_sD_s^+$, 
$D_s^-D^{*+}_s$ and $D^{*-}_sD^{*+}_s$ in the
$\eta$-charmonium dissociation. The $K$-charmonium dissociation includes
9 reactions and the $\eta$-charmonium dissociation includes 18 reactions.
The cross sections for the 27 reactions are calculated with the 
temperature-dependent quark potential, in the Born approximation and in
the quark-interchange mechanism. The cross sections for the
$\bar K$-charmonium dissociation reactions are identical to the cross sections
for the $K$-charmonium dissociation reactions. The numerical cross sections 
are parametrized as functions of $\sqrt s$.
The temperature dependence of the peak cross sections of the endothermic
$K$-charmonium dissociation reactions is closely related to the temperature
dependence of the $K$ radius, the charmonia radii, the confinement
and the node of the $\psi'$ wave function. For $0.6 \leq T/T_{\rm c}<1$ the
cross sections for $\bar{D}^*D$, $\bar{D}D^*$ and $\bar{D}^*D^*$ 
produced in the endothermic $\eta+J/\psi$ reactions around their 
cross-section peaks are larger than the ones for $\bar{D}^*D^+_s$, 
$\bar{D}D^{*+}_s$ and $\bar{D}^*D^{*+}_s$ produced in the endothermic
$K+J/\psi$ reactions, respectively. For $0.6 \leq T/T_{\rm c}<1$ the
cross sections for the $\eta$-charmonium dissociation reactions are
larger than the cross sections for the $\pi$-charmonium dissociation reactions.
To the $\psi'$ and $\chi_c$ suppression in hadronic matter the $\eta +\psi'$
and $\eta + \chi_c$ dissociation reactions may be neglected, but to the 
$J/\psi$ suppression the $\eta + J/\psi$ dissociation reactions need to be 
taken into account.

\section*{Acknowledgements}

This work was supported by the National Natural Science Foundation of China 
under Grant No. 11175111. We thank P. McGuire and H. J. Weber for careful 
readings of the manuscript.

\vspace{0.5cm}
\leftline{\bf Appendix A}
\vspace{0.5cm}

The numerical cross sections for the endothermic reactions in Figs. 3-13 and 
18-23 can be parametrized as
\begin{eqnarray*}
\sigma^{\rm unpol}(\sqrt {s},T)
&=&a_1 \left( \frac {\sqrt {s} -\sqrt {s_0}} {b_1} \right)^{c_1}
\exp \left[ c_1 \left( 1-\frac {\sqrt {s} -\sqrt {s_0}} {b_1} \right) \right] 
\nonumber \\
&&+ a_2 \left( \frac {\sqrt {s} -\sqrt {s_0}} {b_2} \right)^{c_2}
\exp \left[ c_2 \left( 1-\frac {\sqrt {s} -\sqrt {s_0}} {b_2} \right) \right],
\nonumber \\
\end{eqnarray*}
and for the exothermic reactions in Figs. 6-10, 12-17 and 20-21
\begin{eqnarray*}
\sigma^{\rm unpol}(\sqrt {s},T)
&=&\frac{\vec{P}^{\prime 2}}{\vec{P}^2}
\left\{a_1 \left( \frac {\sqrt {s} -\sqrt {s_0}} {b_1} \right)^{c_1}
\exp \left[ c_1 \left( 1-\frac {\sqrt {s} -\sqrt {s_0}} {b_1} \right) \right] 
\right.
\nonumber \\
&&+ \left.
a_2 \left( \frac {\sqrt {s} -\sqrt {s_0}} {b_2} \right)^{c_2}
\exp \left[ c_2 \left( 1-\frac {\sqrt {s} -\sqrt {s_0}} {b_2} \right) \right]
\right\},
\nonumber \\
\end{eqnarray*}
with
\begin{eqnarray*}
\vec {P}^2(\sqrt{s})=\frac{1}{4s}\left\{ \left[ s-\left (
m_{q\bar {q}}^{2}+m_{c\bar {c}}^{2}\right) \right]^{2}
-4m_{q\bar {q}}^{2}m_{c\bar {c}}^{2}\right\}, 
\nonumber \\
\vec {P'}^{2}(\sqrt{s})=\frac{1}{4s}\left\{ \left [
s-\left( m_{q\bar {c}}^2+m_ {c\bar{q}}^2\right)\right]^2
-4m_ {q\bar {c}} ^2m_ {c\bar{q}}^2 \right\}.
\nonumber \\
\end{eqnarray*}
Here $a_1$, $b_1$, $c_1$, $a_2$, $b_2$ and $c_2$ are parameters and 
$\sqrt{s_0}$ is the threshold energy. The parameter values are listed in Tables
1-6. We also list $d_0$ what is the separation between the peak's location on 
the $\sqrt s$-axis and the threshold energy and $\sqrt{s_z}$ which is 
the square root of the Mandelstam variable at which the cross section is 
1/100 of the peak cross section. According to a procedure presented in Ref. 
\cite{ZX} we can get cross sections at any temperature from the quantities
listed in Tables 1-6.


\newpage
\begin{figure}[htbp]
  \centering
    \includegraphics[width=65mm,height=60mm,angle=0]{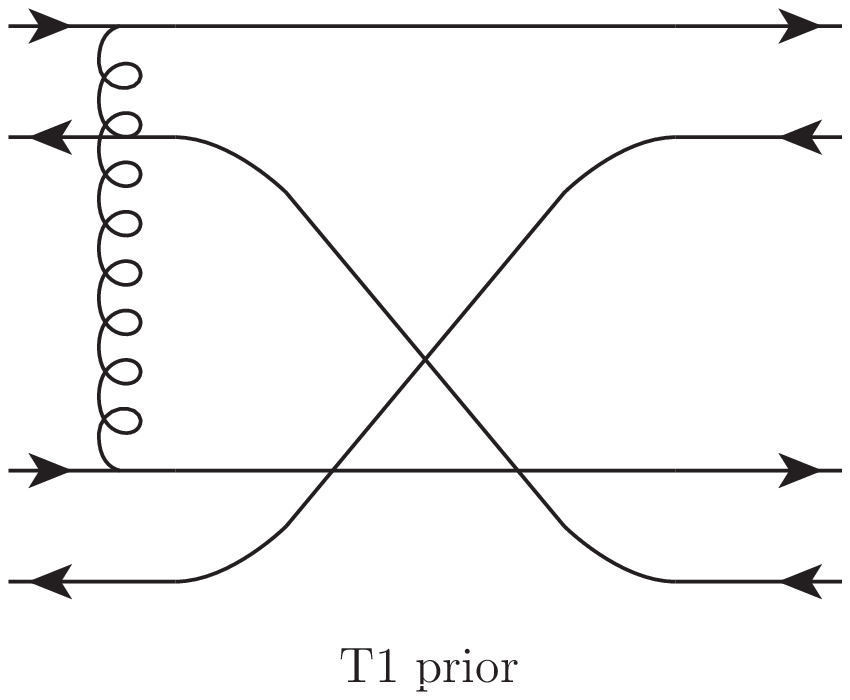}
      \hspace{1.5cm}
    \includegraphics[width=65mm,height=60mm,angle=0]{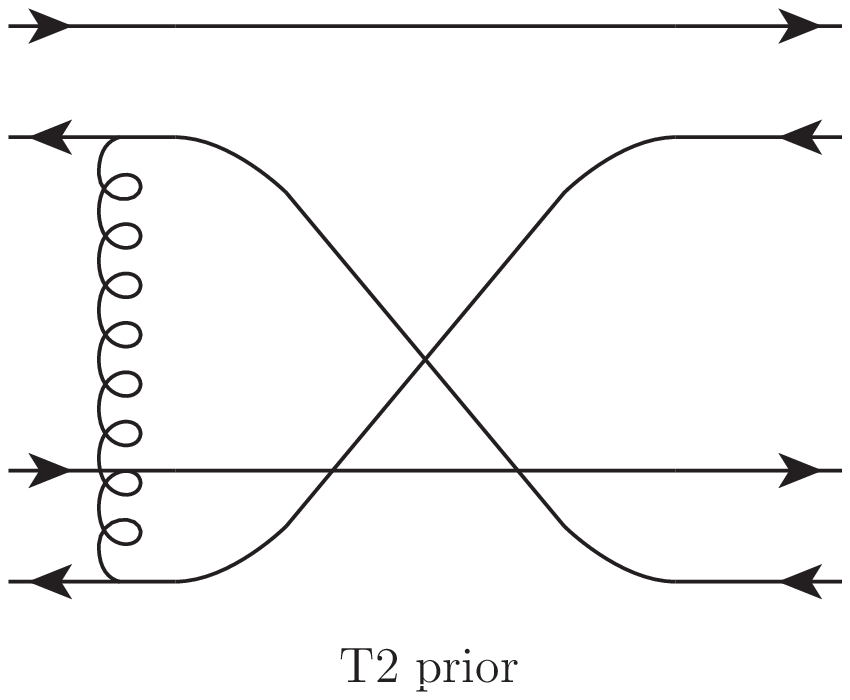}
      \vskip 36pt
    \includegraphics[width=65mm,height=60mm,angle=0]{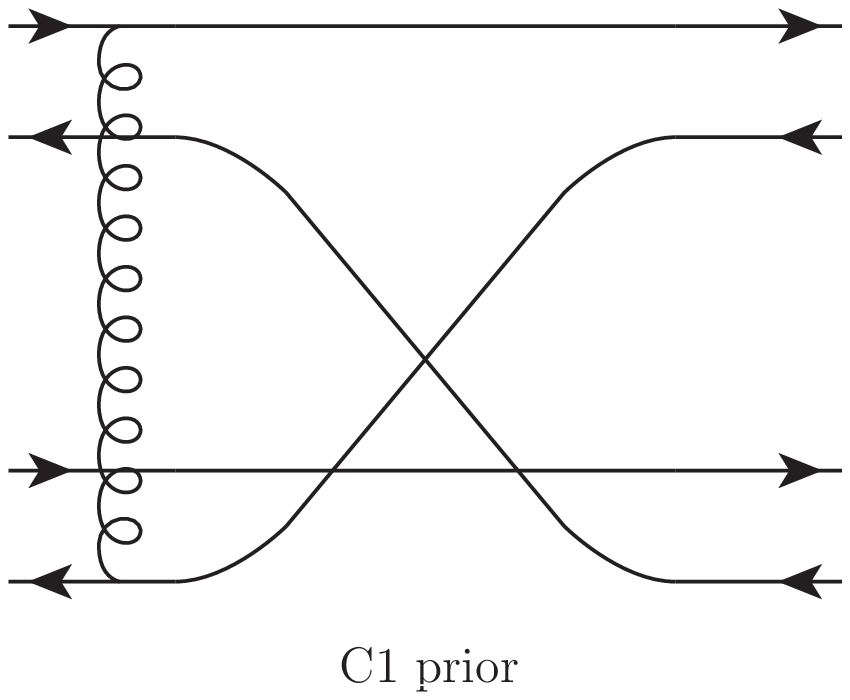}
      \hspace{1.5cm}
    \includegraphics[width=65mm,height=60mm,angle=0]{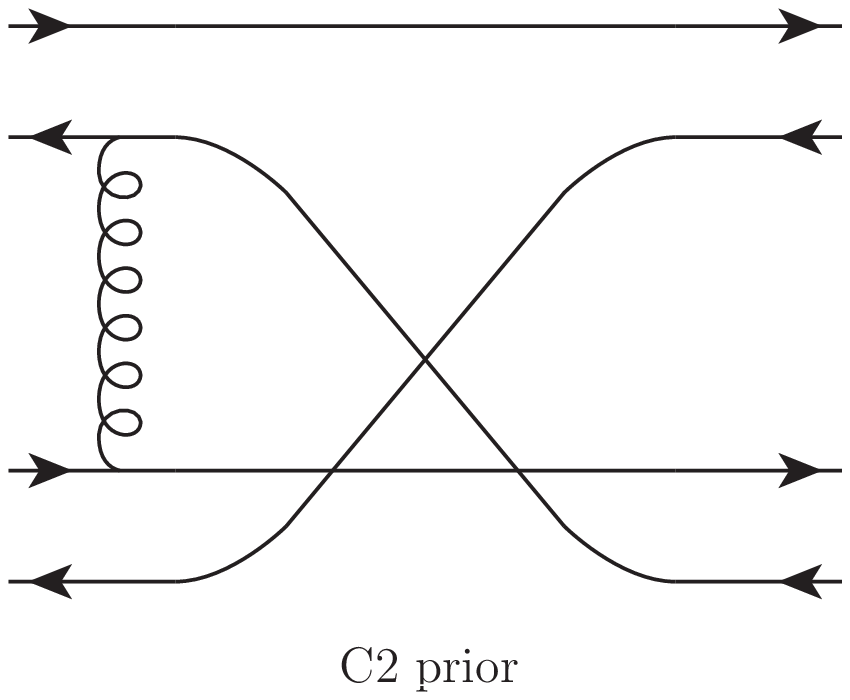}
\caption{'Prior' diagrams. Solid (wavy) lines represent quarks or antiquarks
(gluons).}
\label{fig1}
\end{figure}

\newpage
\begin{figure}[htbp]
  \centering
    \includegraphics[width=65mm,height=60mm,angle=0]{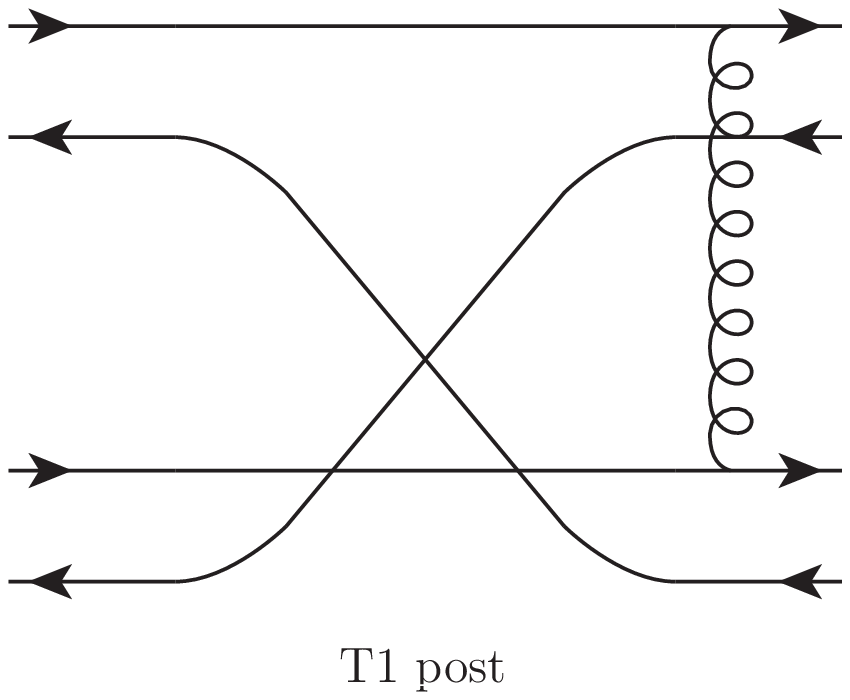}
      \hspace{1.5cm}
    \includegraphics[width=65mm,height=60mm,angle=0]{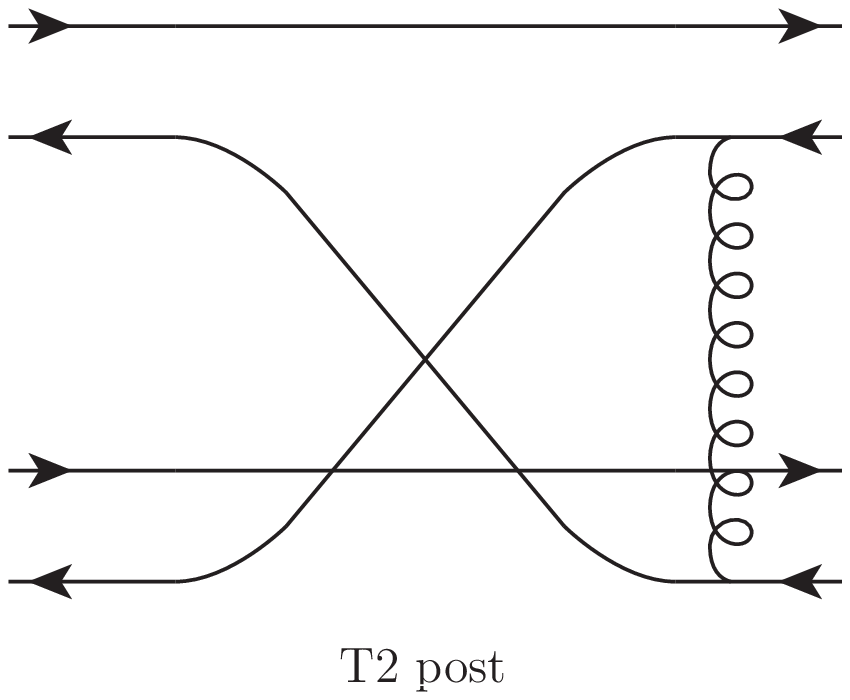}
      \vskip 36pt
    \includegraphics[width=65mm,height=60mm,angle=0]{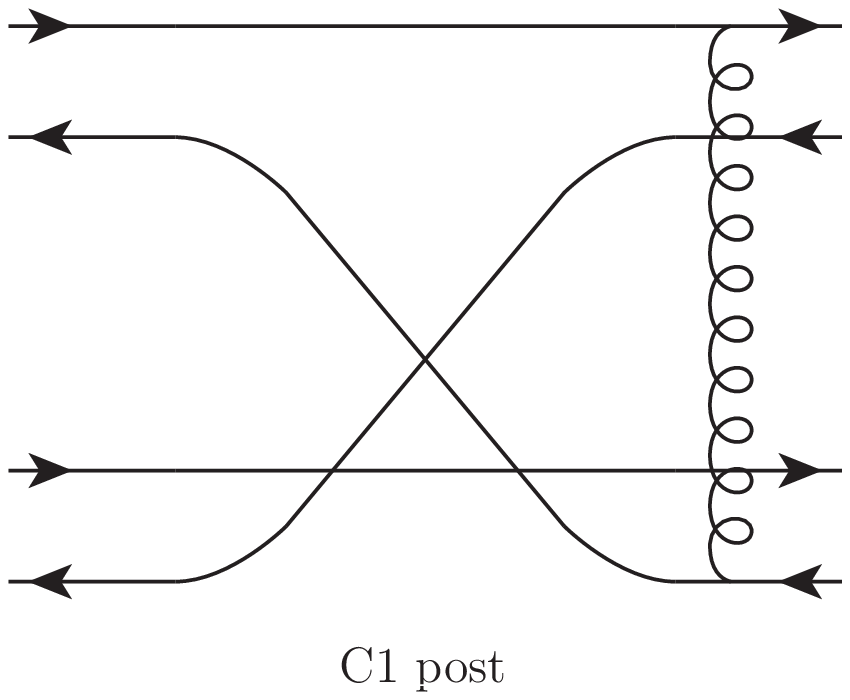}
      \hspace{1.5cm}
    \includegraphics[width=65mm,height=60mm,angle=0]{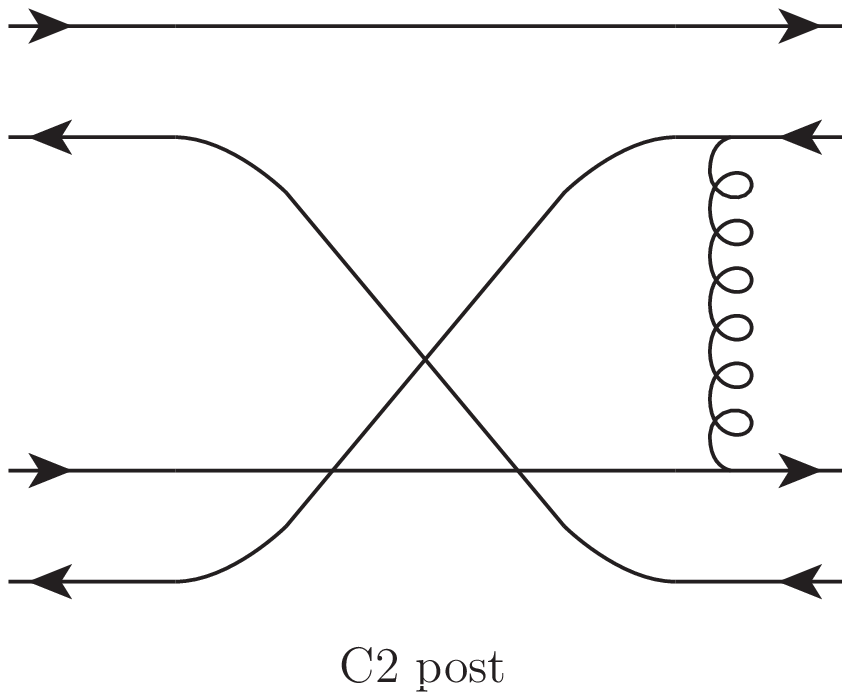}
\caption{'Post' diagrams. Solid (wavy) lines represent quarks or antiquarks
(gluons).}
\label{fig2}
\end{figure}

\newpage
\begin{figure}[htbp]
\centering
\includegraphics[scale=0.65]{kjpsidadsrltvt.eps}
\caption{Cross sections for $K+J/\psi\to\bar{D}^{*}+D^+_s$ at
various temperatures.}
\label{fig3}
\end{figure}

\newpage
\begin{figure}[htbp]
\centering
\includegraphics[scale=0.65]{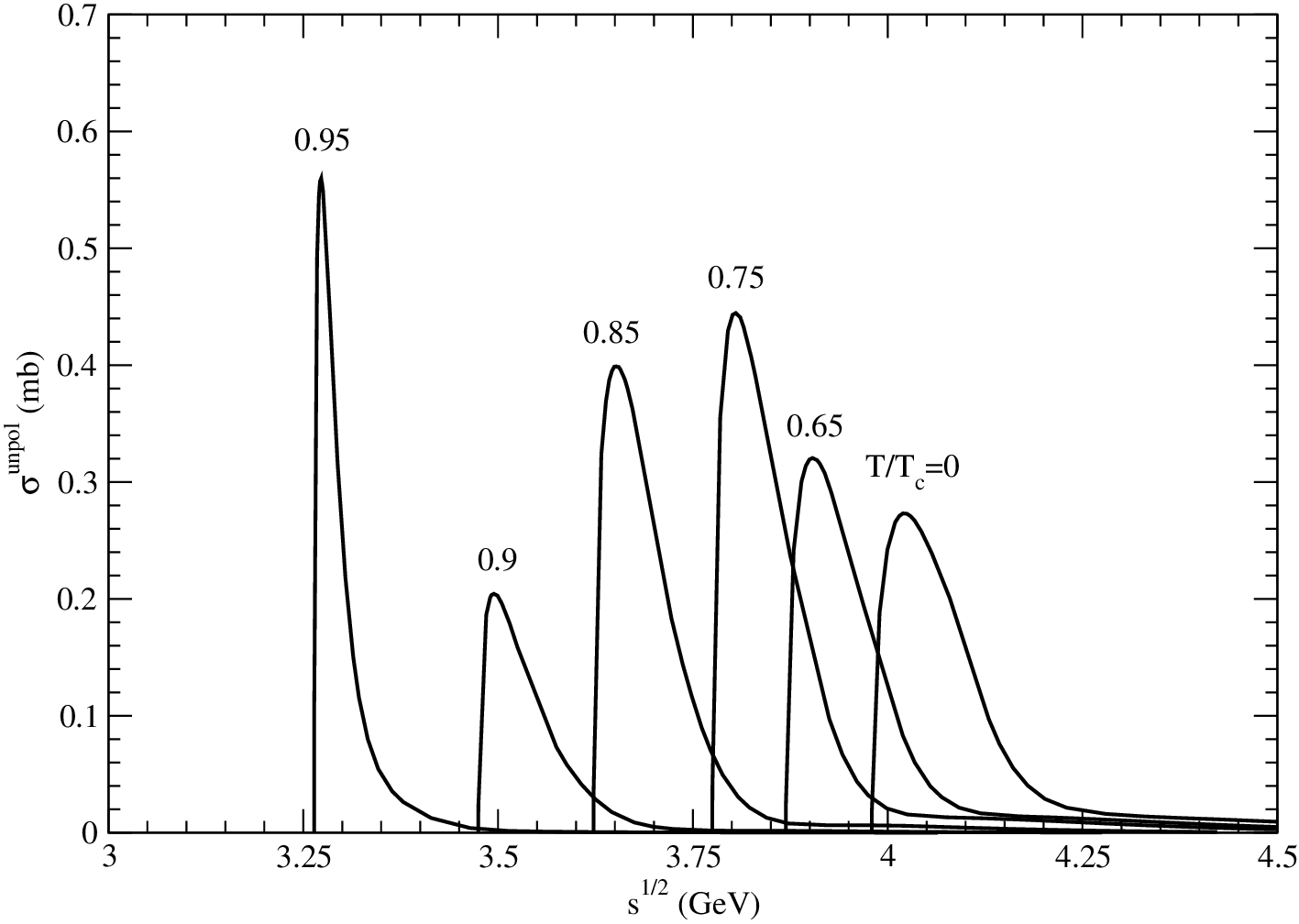}
\caption{Cross sections for $K+J/\psi\to\bar{D}+D^{*+}_s$ at
various temperatures.}
\label{fig4}
\end{figure}

\newpage
\begin{figure}[htbp]
\centering
\includegraphics[scale=0.64]{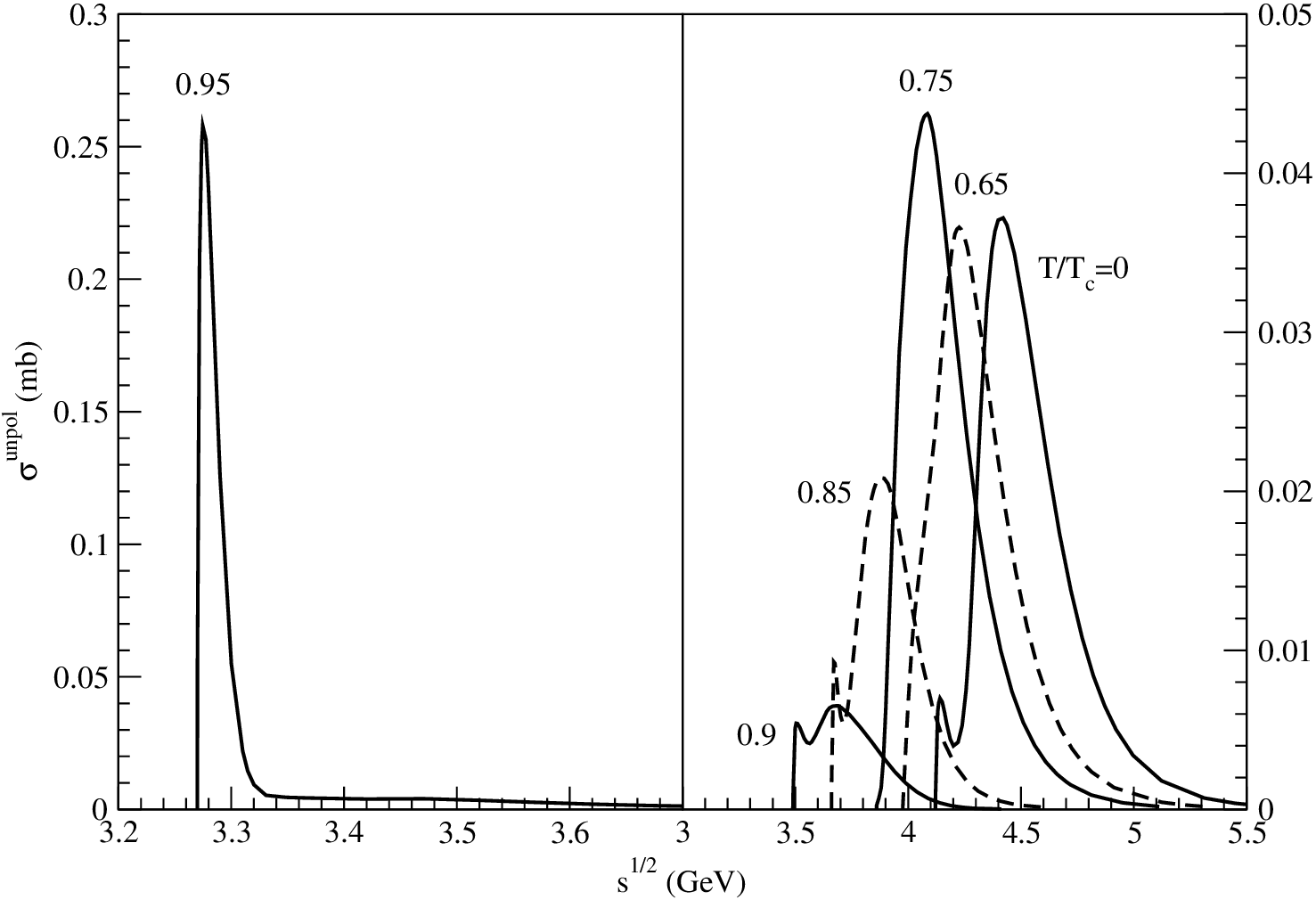}
\caption{Cross sections for $K+J/\psi\to\bar{D}^{*}+D^{*+}_s$ at
various temperatures.}
\label{fig5}
\end{figure}

\newpage
\begin{figure}[htbp]
\centering
\includegraphics[scale=0.65]{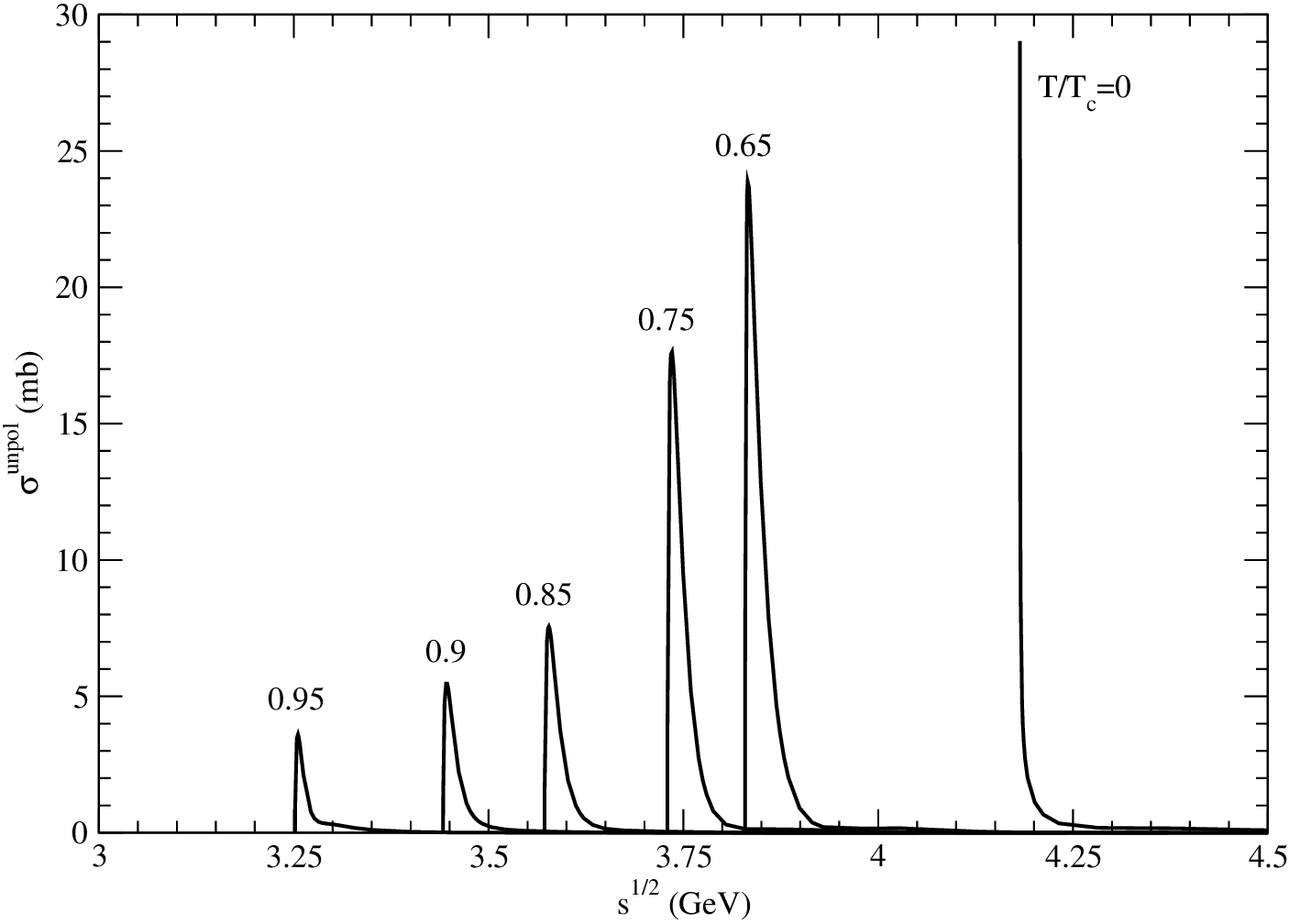}
\caption{Cross sections for $K+\psi'\to\bar{D}^{*}+D^+_s$ at
various temperatures.}
\label{fig6}
\end{figure}

\newpage
\begin{figure}[htbp]
\centering
\includegraphics[scale=0.65]{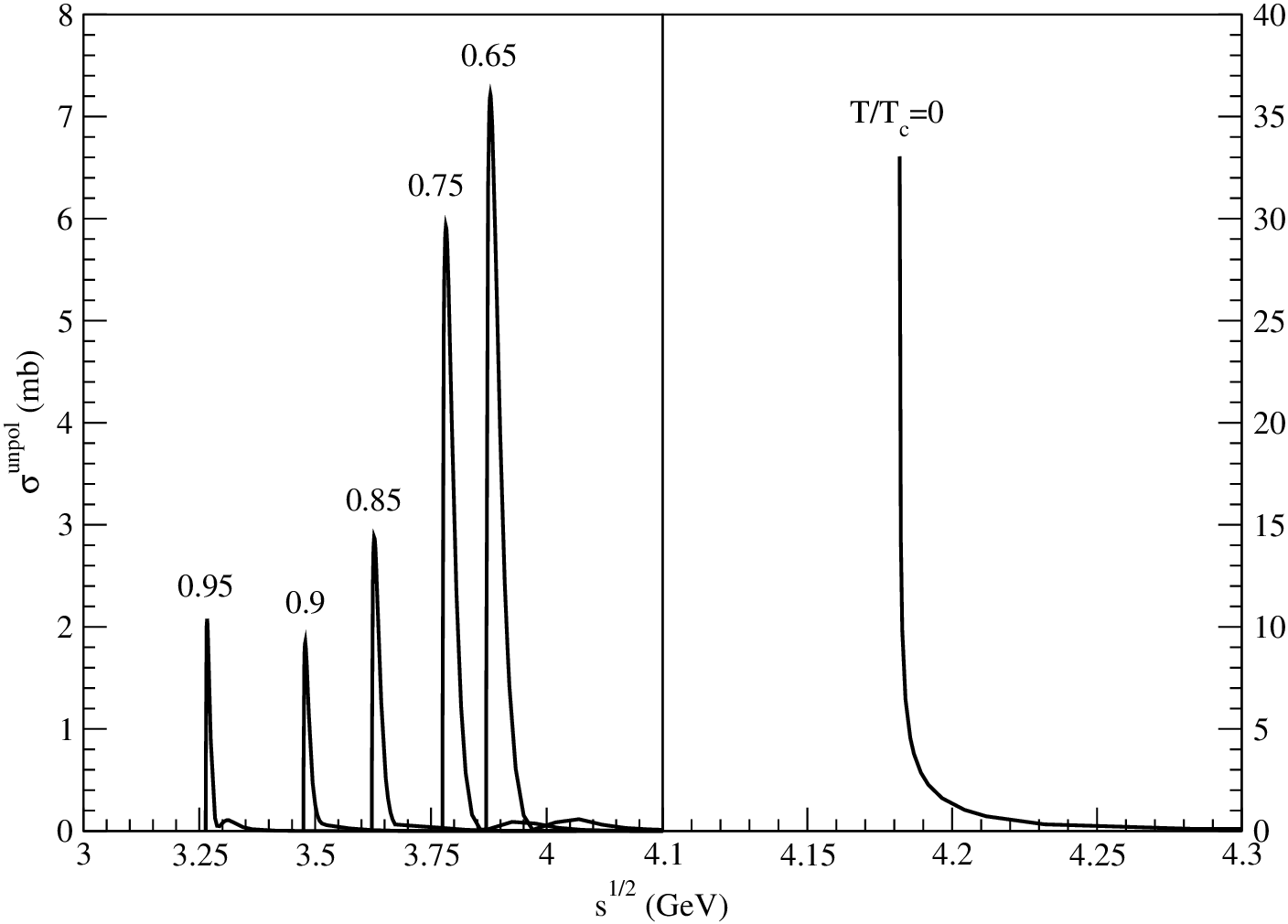}
\caption{Cross sections for $K+\psi'\to\bar{D}+D^{*+}_s$ at
various temperatures.}
\label{fig7}
\end{figure}

\newpage
\begin{figure}[htbp]
\centering
\includegraphics[scale=0.65]{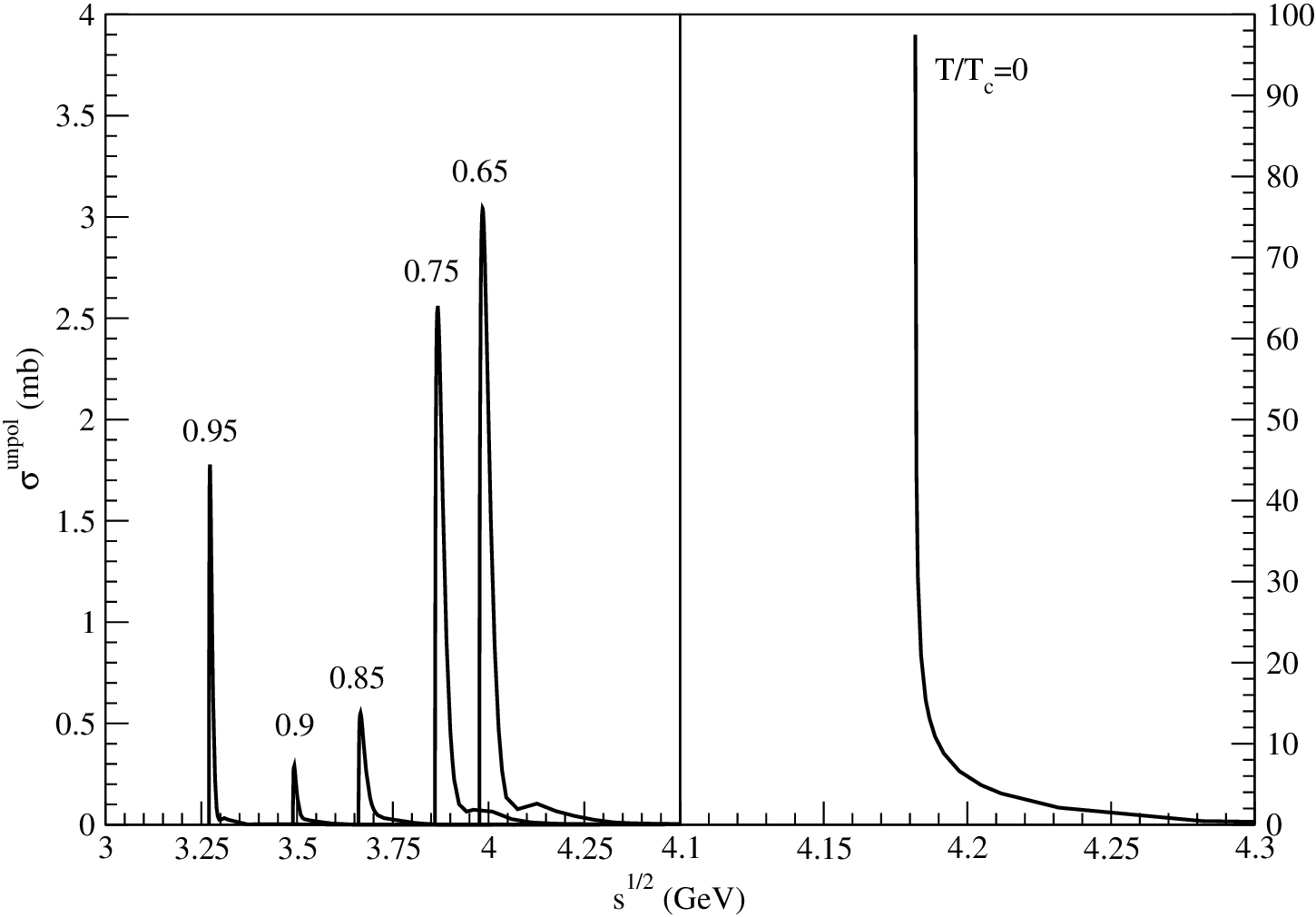}
\caption{Cross sections for $K+\psi'\to\bar{D}^{*}+D^{*+}_s$ at
various temperatures.}
\label{fig8}
\end{figure}

\newpage
\begin{figure}[htbp]
\centering
\includegraphics[scale=0.65]{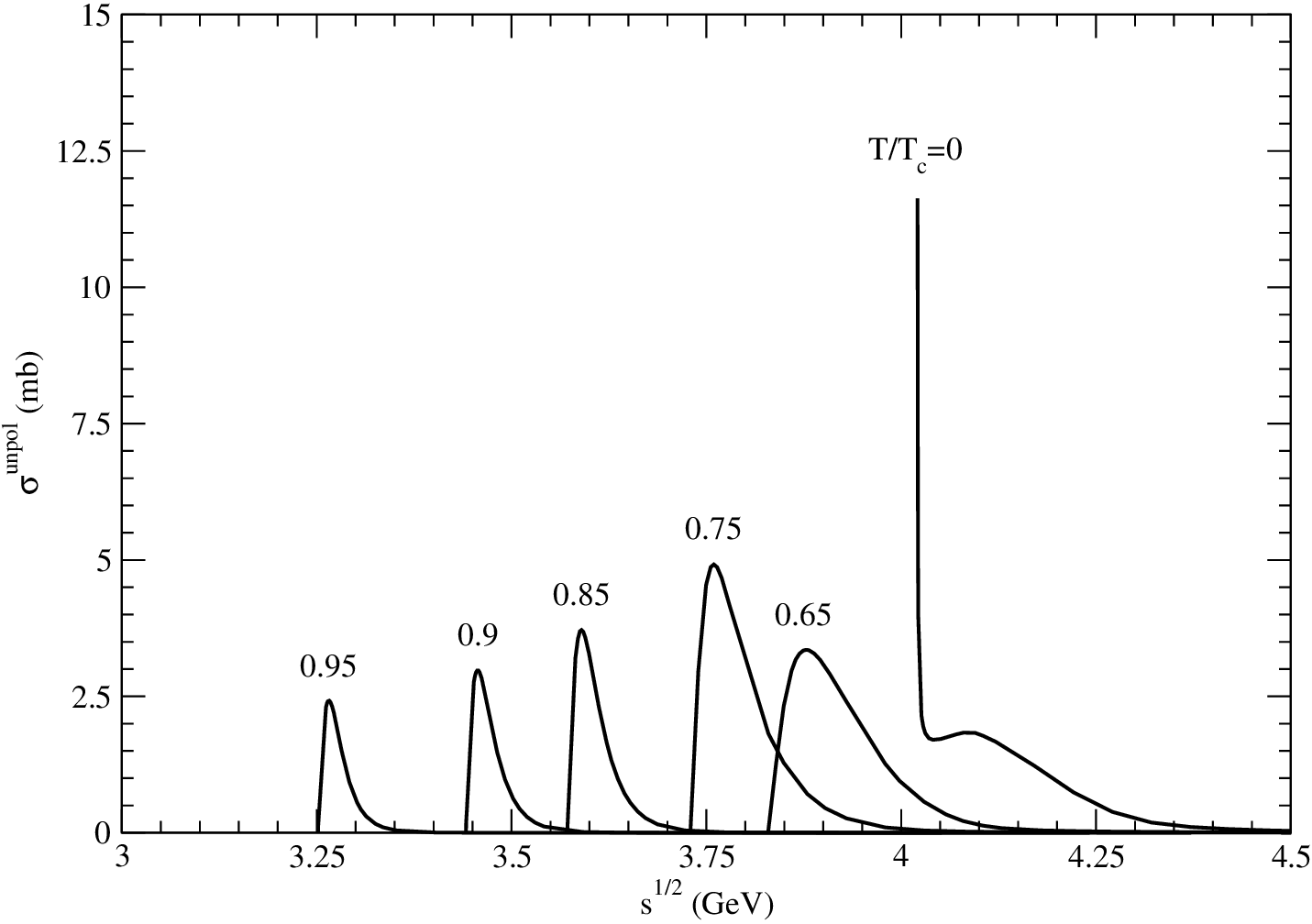}
\caption{Cross sections for $K+\chi_{c}\to\bar{D}^{*}+D^+_s$ at
various temperatures.}
\label{fig9}
\end{figure}

\newpage
\begin{figure}[htbp]
\centering
\includegraphics[scale=0.65]{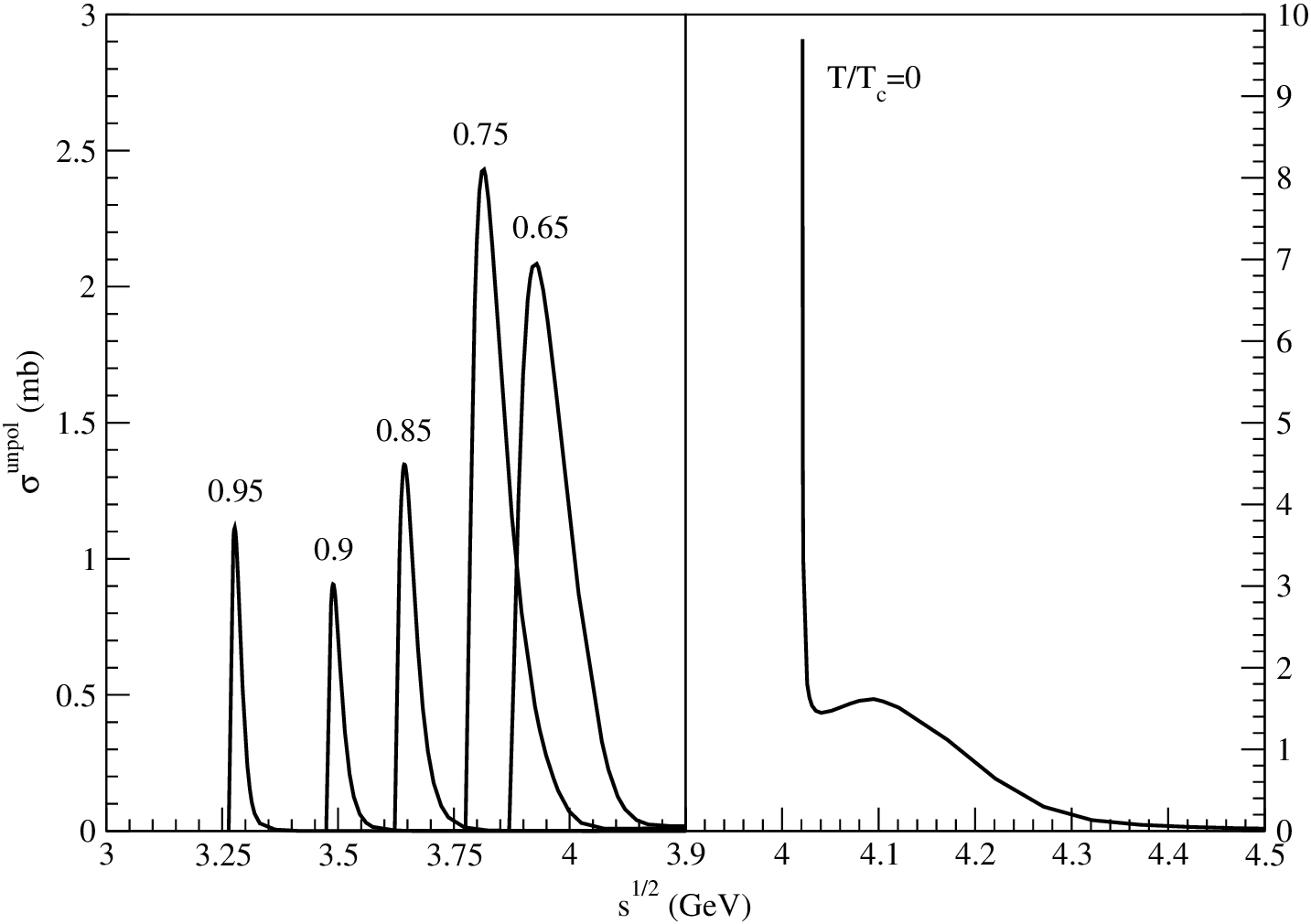}
\caption{Cross sections for $K+\chi_{c}\to\bar{D}+D^{*+}_s$ at
various temperatures.}
\label{fig10}
\end{figure}

\newpage
\begin{figure}[htbp]
\centering
\includegraphics[scale=0.65]{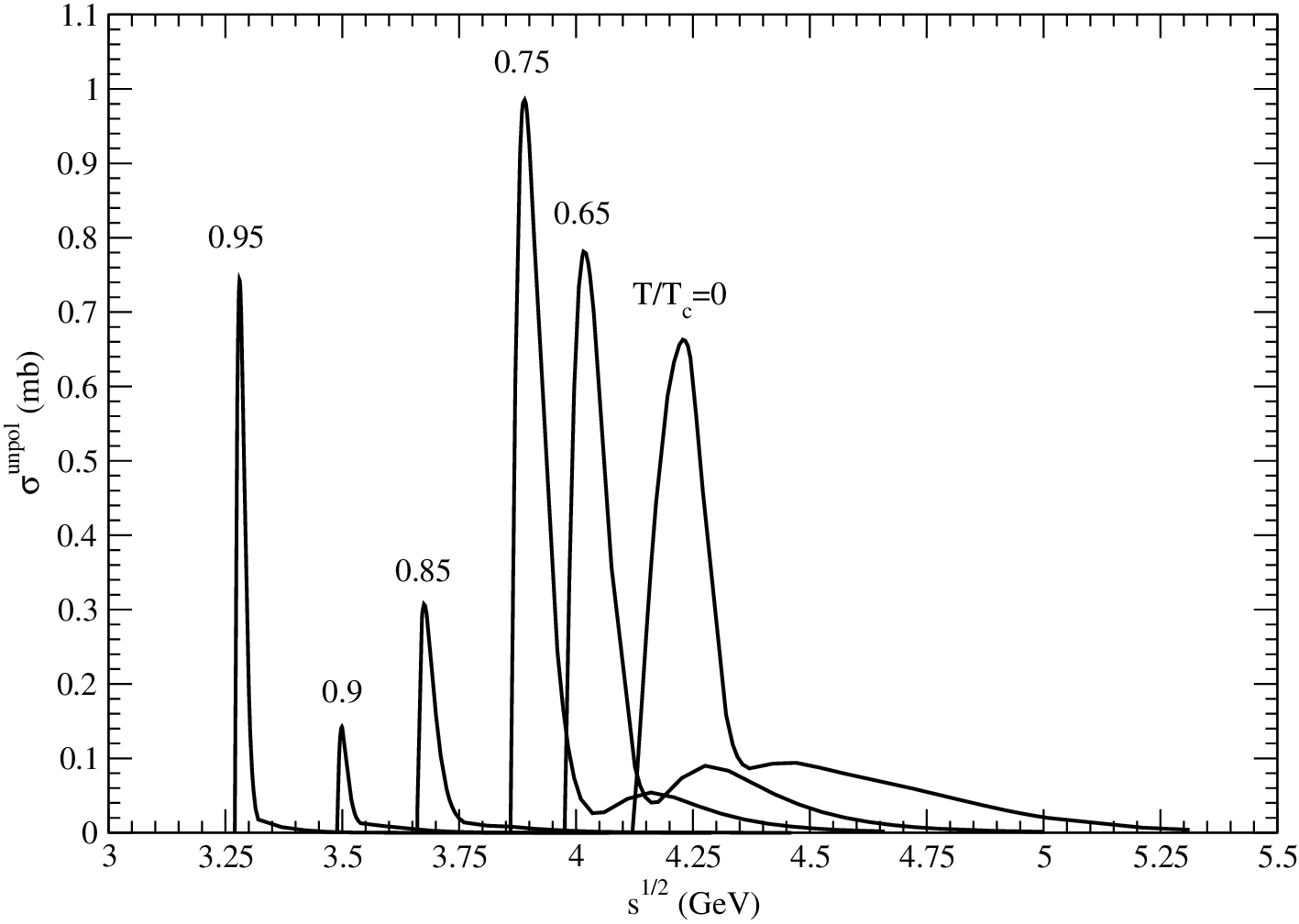}
\caption{Cross sections for $K+\chi_{c}\to\bar{D}^{*}+D^{*+}_s$ at
various temperatures.}
\label{fig11}
\end{figure}

\newpage
\begin{figure}[htbp]
\centering
\includegraphics[scale=0.65]{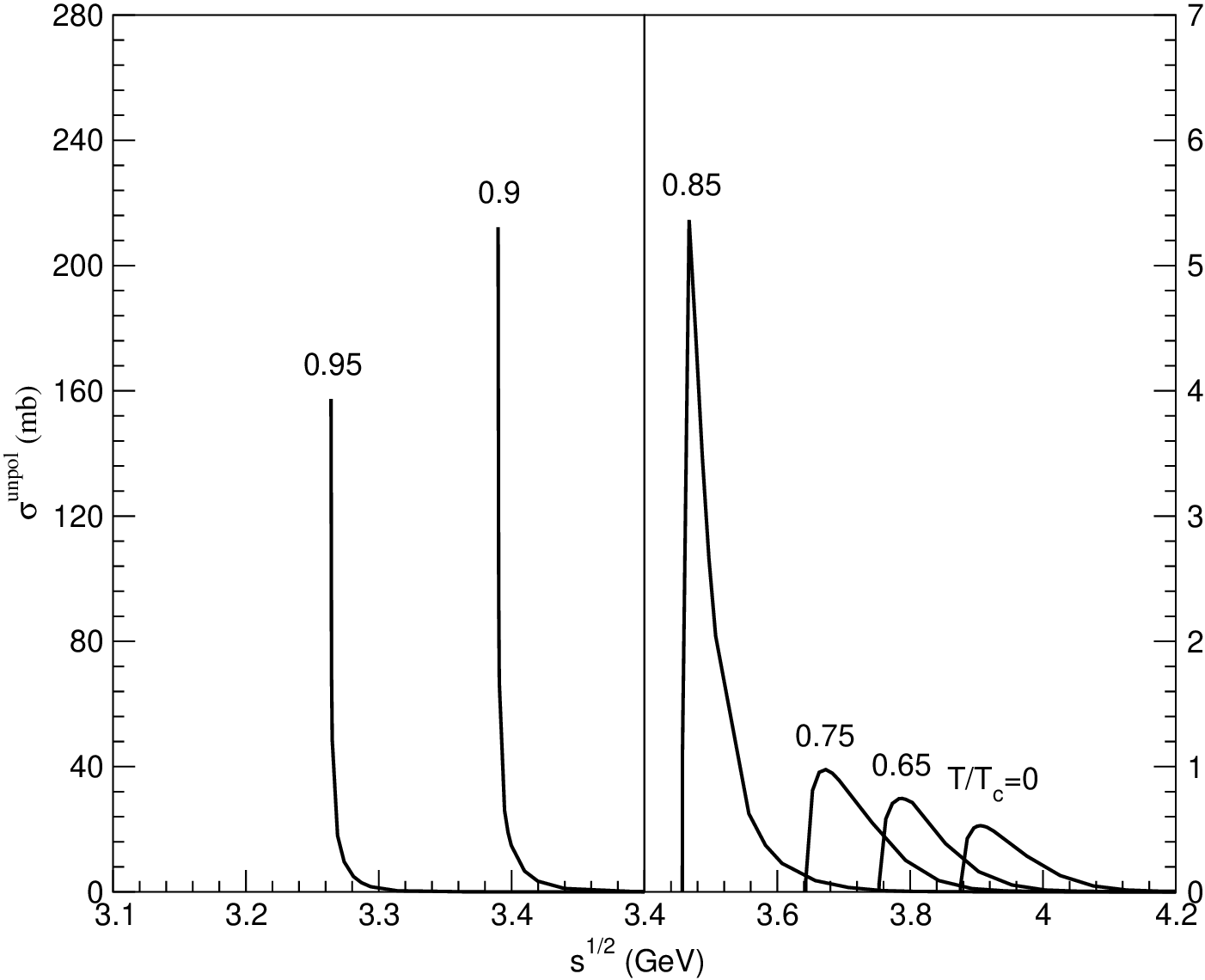}
\caption{Cross sections for $\eta+J/\psi\to\bar{D}^{*}+D$ or $\bar{D}+D^{*}$ at
various temperatures.}
\label{fig12}
\end{figure}

\newpage
\begin{figure}[htbp]
\centering
\includegraphics[scale=0.62]{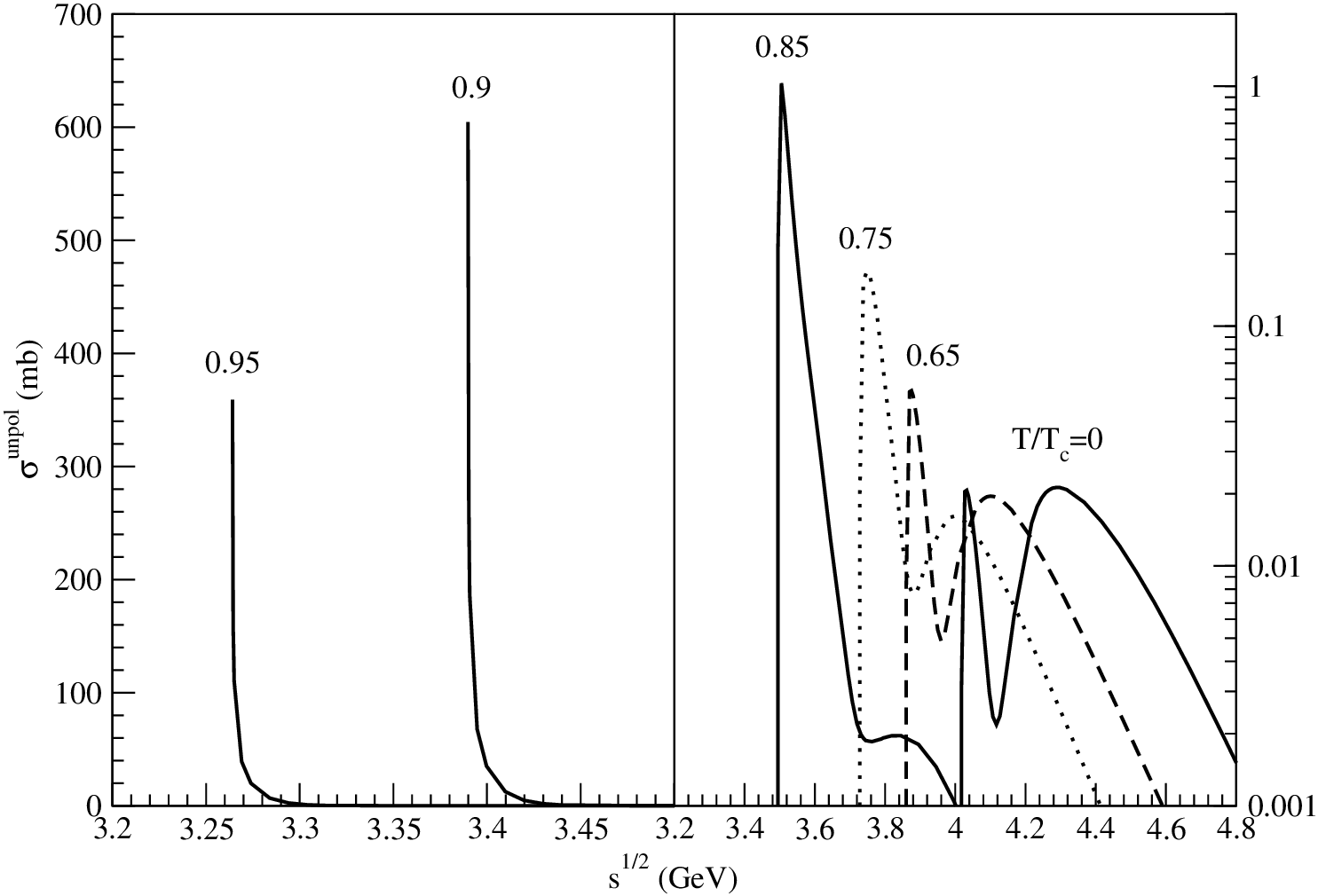}
\caption{Cross sections for $\eta+J/\psi\to\bar{D}^{*}+D^{*}$ at
various temperatures.}
\label{fig13}
\end{figure}

\newpage
\begin{figure}[htbp]
\centering
\includegraphics[scale=0.65]{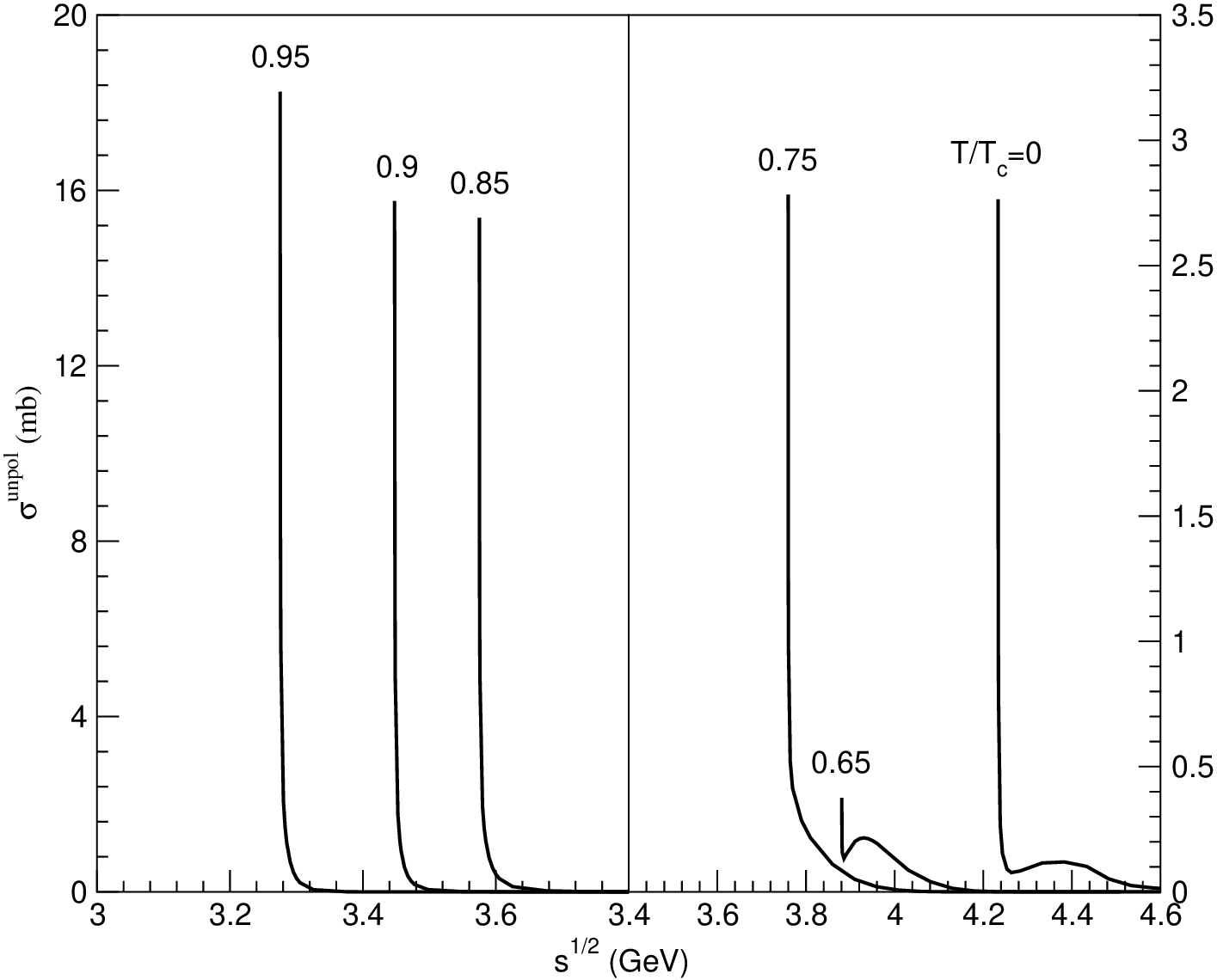}
\caption{Cross sections for $\eta+\psi'\to\bar{D}^{*}+D$ or $\bar{D}+D^{*}$ at
various temperatures.}
\label{fig14}
\end{figure}

\newpage
\begin{figure}[htbp]
\centering
\includegraphics[scale=0.65]{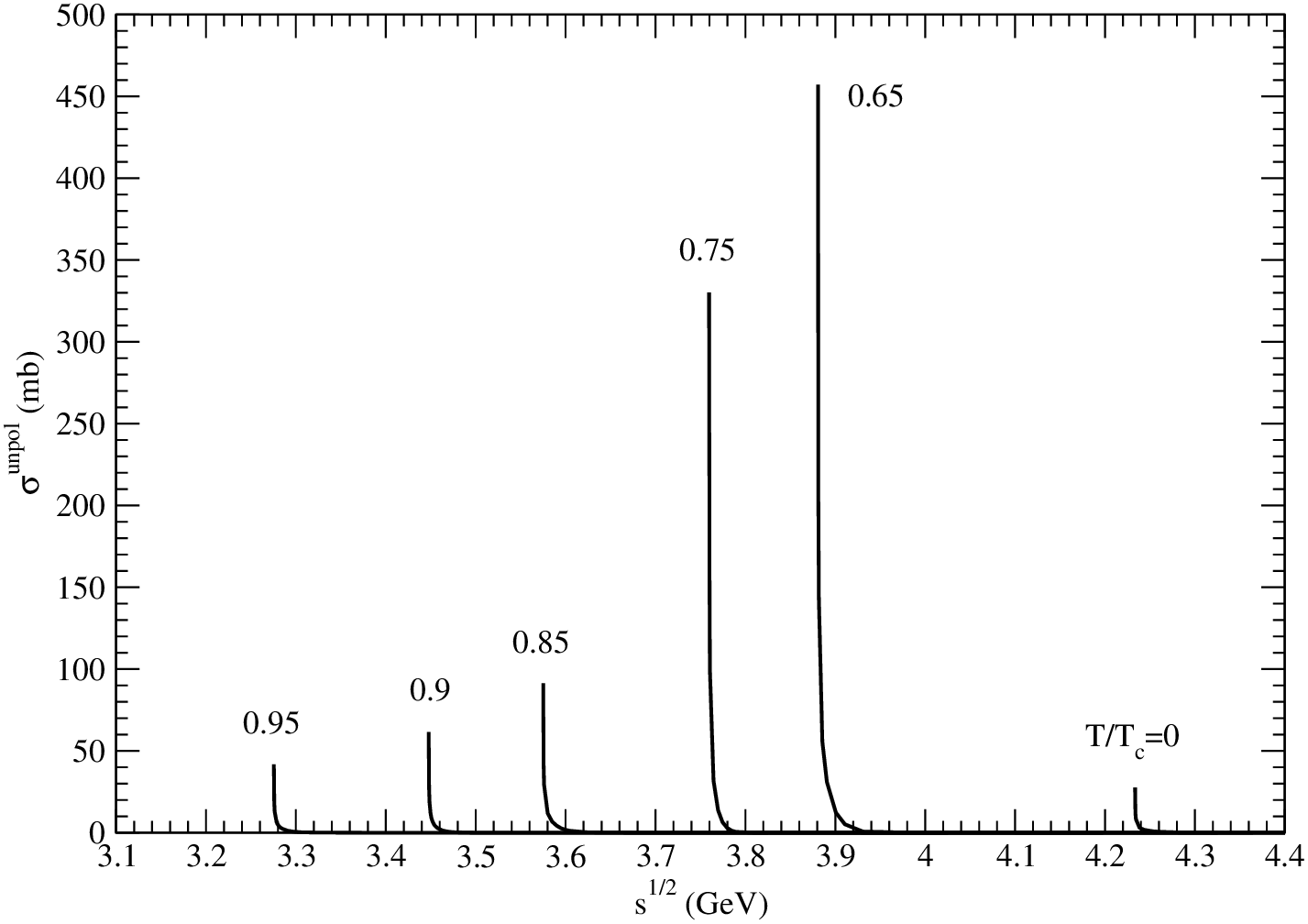}
\caption{Cross sections for $\eta+\psi'\to\bar{D}^{*}+D^{*}$ at
various temperatures.}
\label{fig15}
\end{figure}

\newpage
\begin{figure}[htbp]
\centering
\includegraphics[scale=0.62]{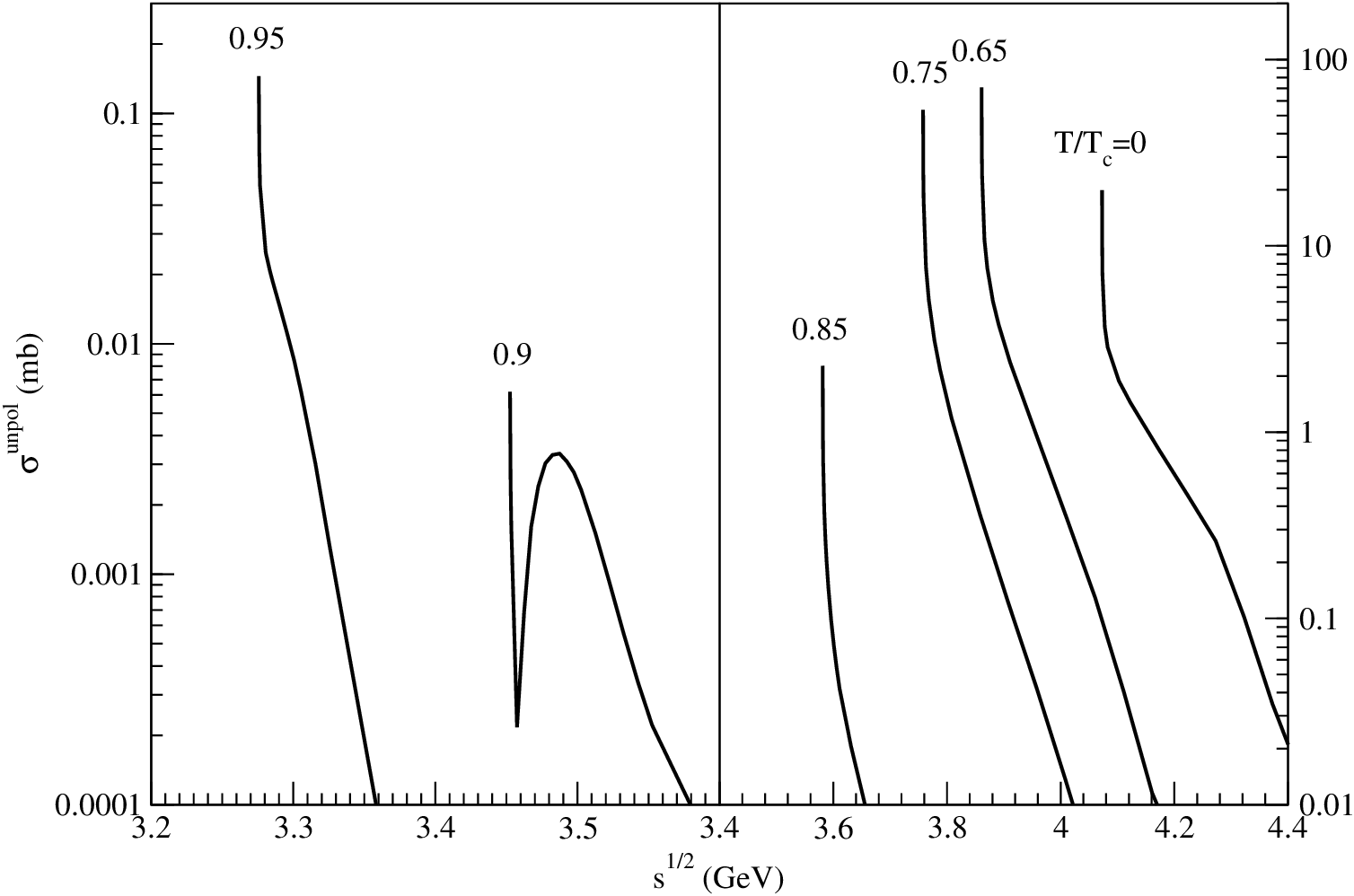}
\caption{Cross sections for $\eta+\chi_{c}\to\bar{D}^{*}+D$ or $\bar{D}+D^{*}$ 
at various temperatures.}
\label{fig16}
\end{figure}

\newpage
\begin{figure}[htbp]
\centering
\includegraphics[scale=0.62]{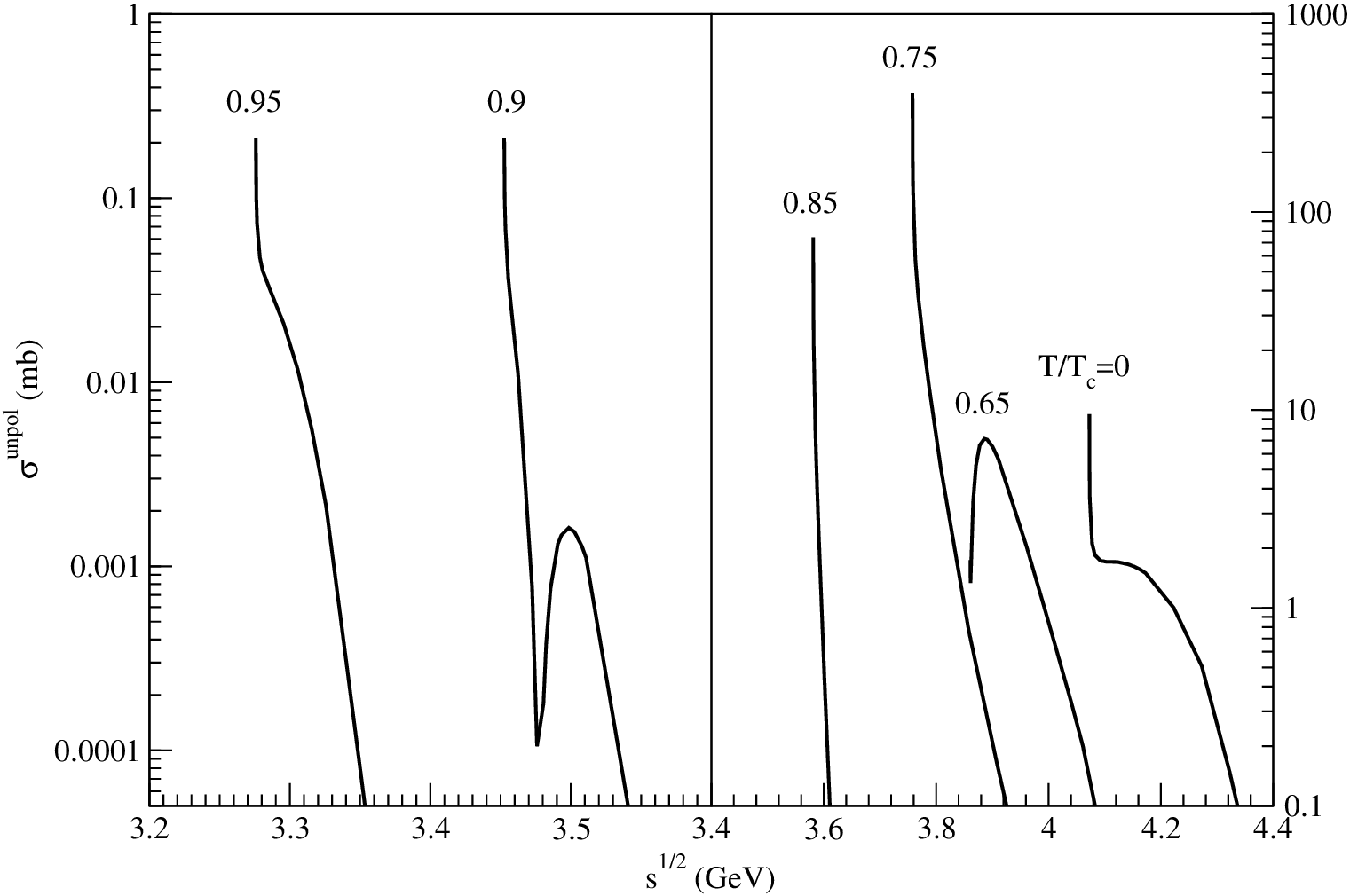}
\caption{Cross sections for $\eta+\chi_{c}\to\bar{D}^{*}+D^{*}$ at
various temperatures.}
\label{fig17}
\end{figure}

\newpage
\begin{figure}[htbp]
\centering
\includegraphics[scale=0.7]{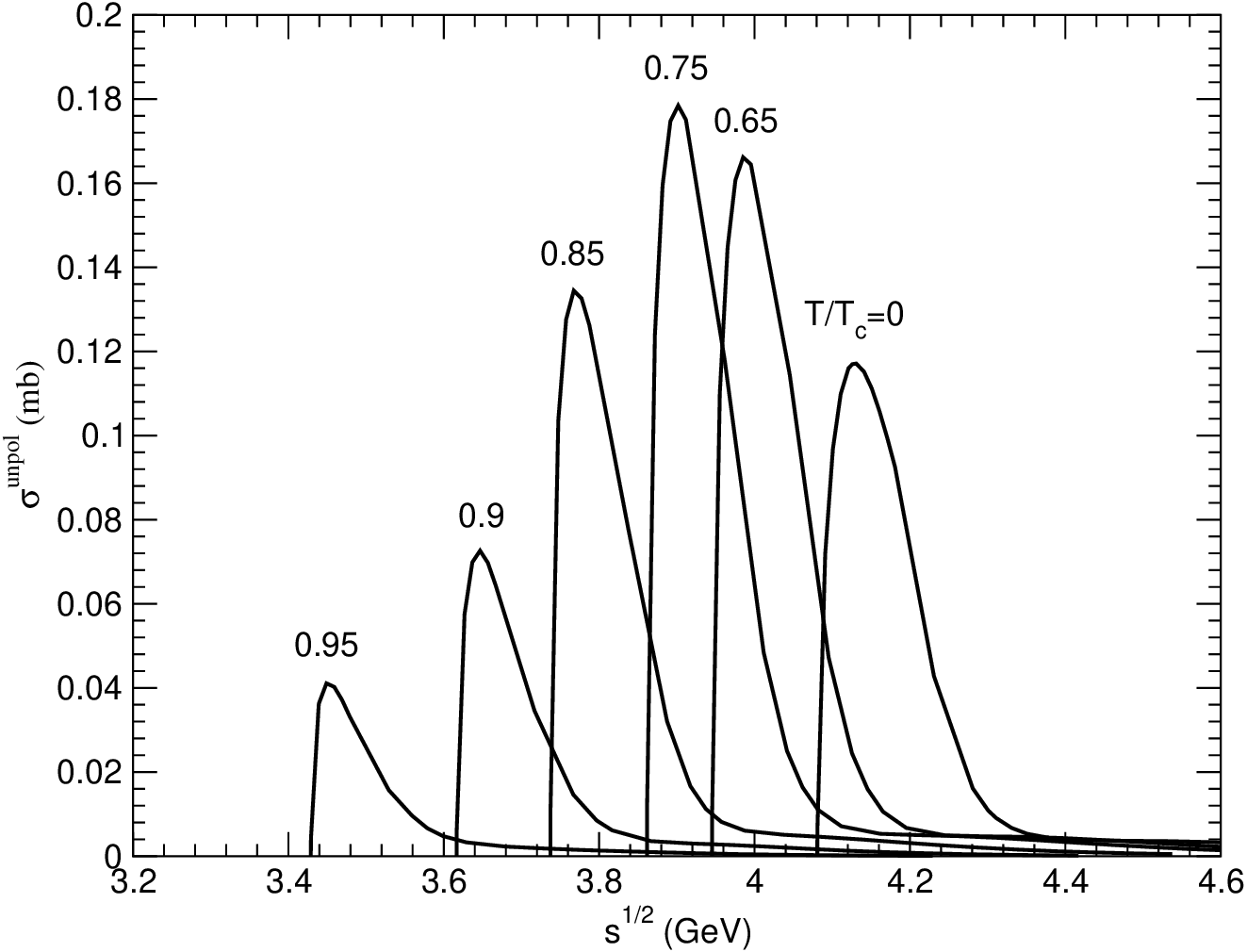}
\caption{Cross sections for $\eta+J/\psi\to D^{*-}_{s}+D_{s}^+$ or 
$D_{s}^-+D^{*+}_{s}$ at various temperatures.}
\label{fig18}
\end{figure}

\newpage
\begin{figure}[htbp]
\centering
\includegraphics[scale=0.7]{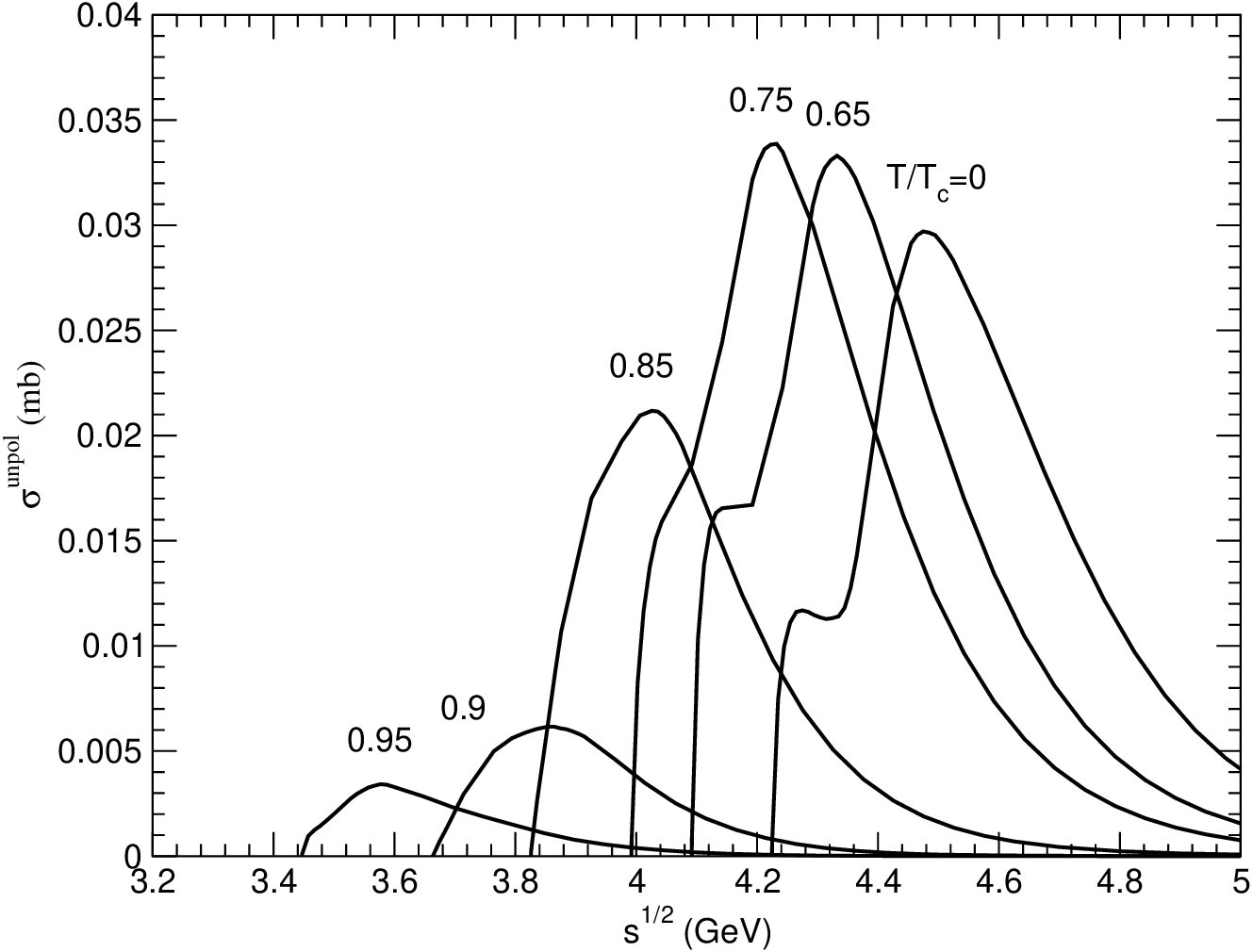}
\caption{Cross sections for $\eta+J/\psi\to D^{*-}_{s}+D^{*+}_{s}$ at
various temperatures.}
\label{fig19}
\end{figure}

\newpage
\begin{figure}[htbp]
\centering
\includegraphics[scale=0.65]{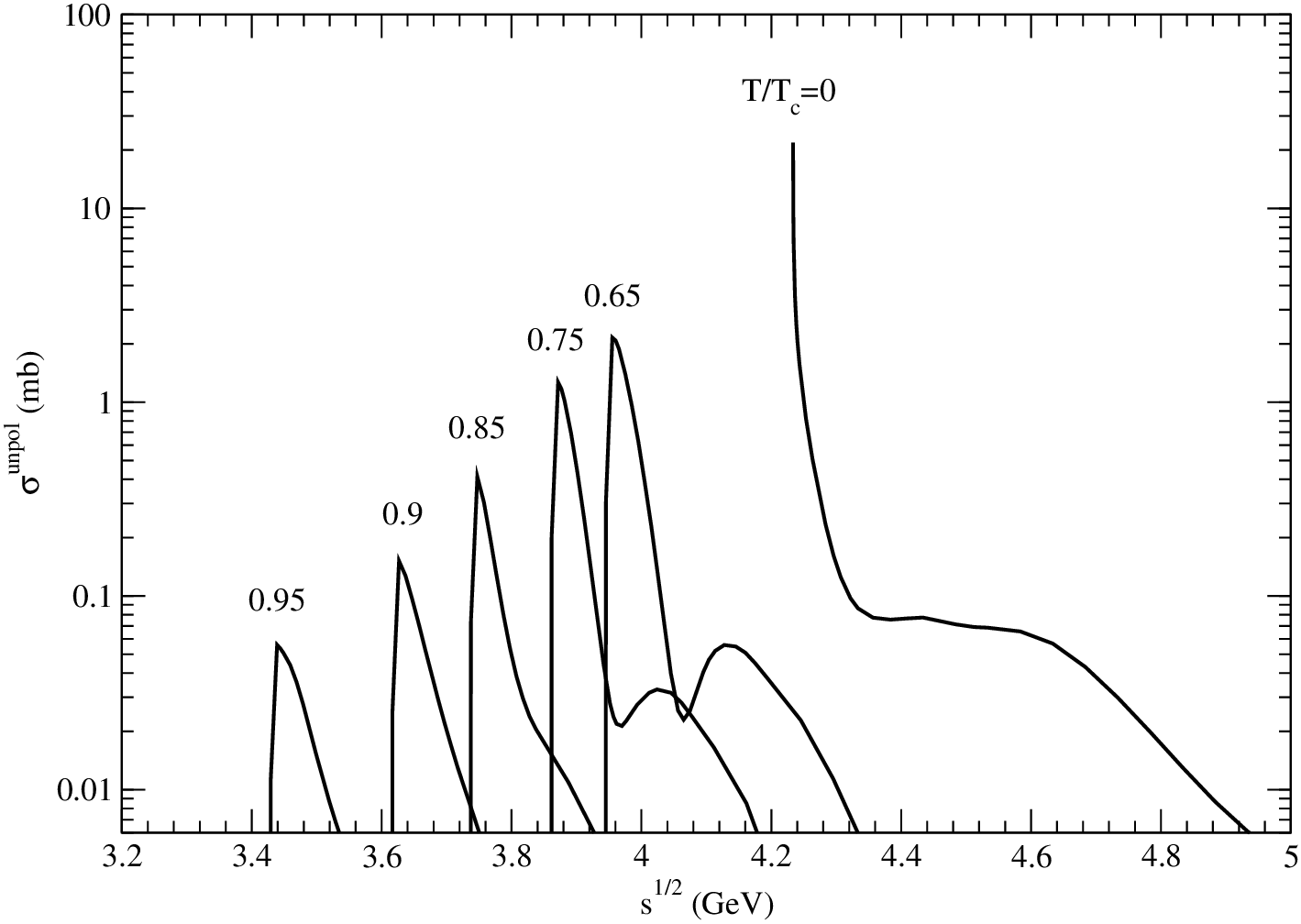}
\caption{Cross sections for $\eta+\psi'\to D^{*-}_{s}+D_{s}^+$ or 
$D_{s}^-+D^{*+}_{s}$ at various temperatures.}
\label{fig20}
\end{figure}

\newpage
\begin{figure}[htbp]
\centering
\includegraphics[scale=0.65]{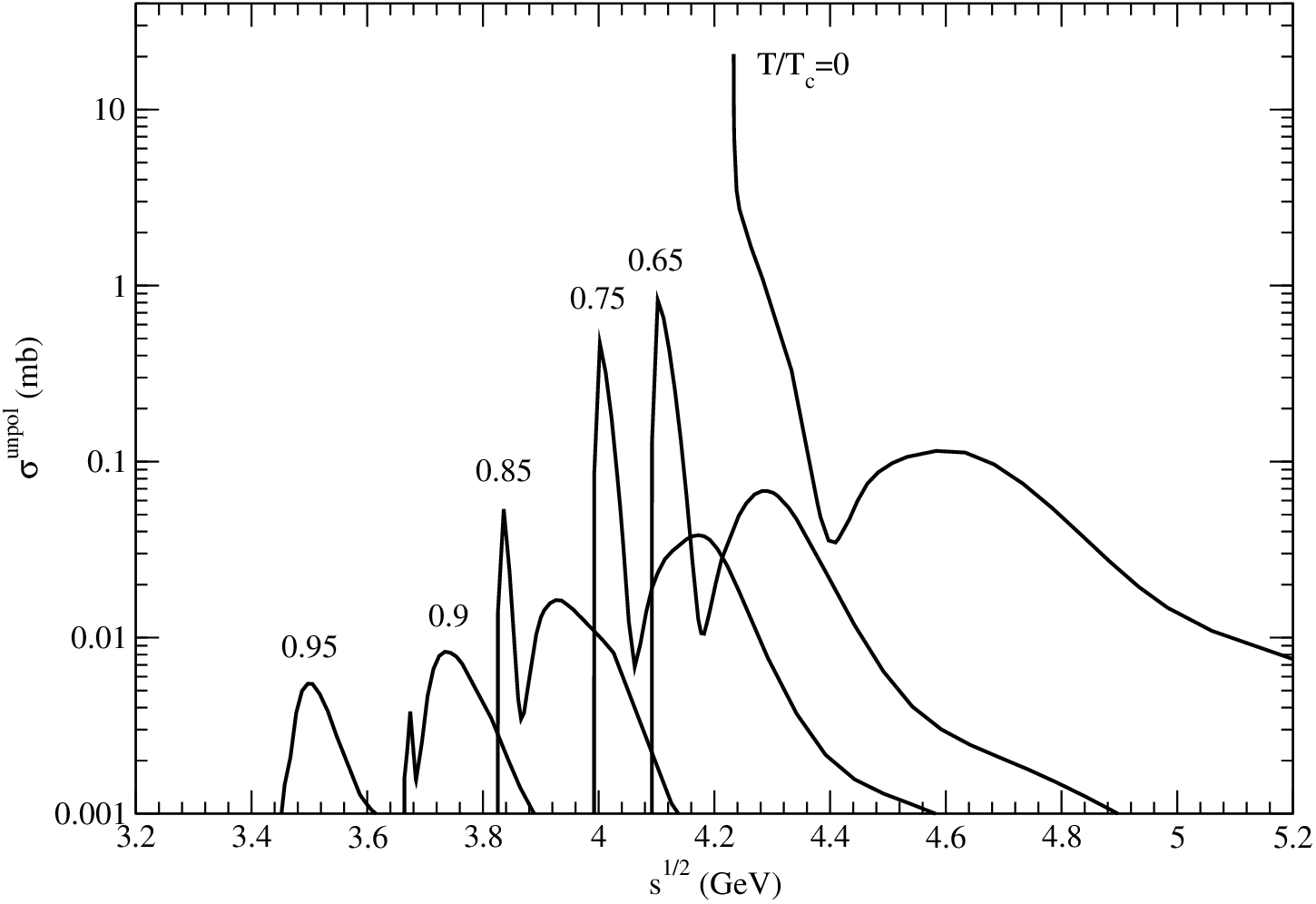}
\caption{Cross sections for $\eta+\psi'\to D^{*-}_{s}+D^{*+}_{s}$ at
various temperatures.}
\label{fig21}
\end{figure}

\newpage
\begin{figure}[htbp]
\centering
\includegraphics[scale=0.7]{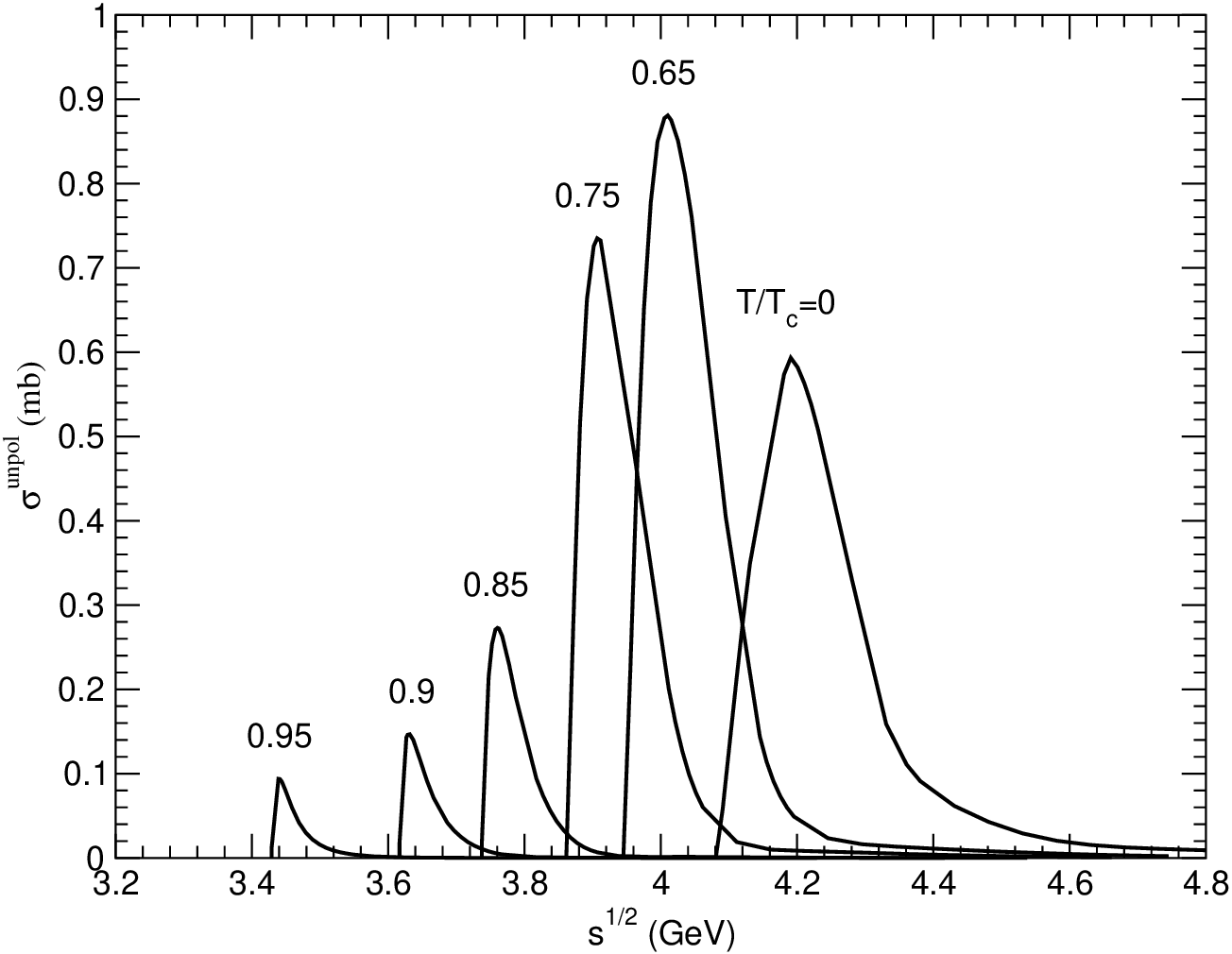}
\caption{Cross sections for $\eta+\chi_{c}\to D^{*-}_{s}+D_{s}^+$ or 
$D_{s}^- + D^{*+}_{s}$ at various temperatures.}
\label{fig22}
\end{figure}

\newpage
\begin{figure}[htbp]
\centering
\includegraphics[scale=0.6]{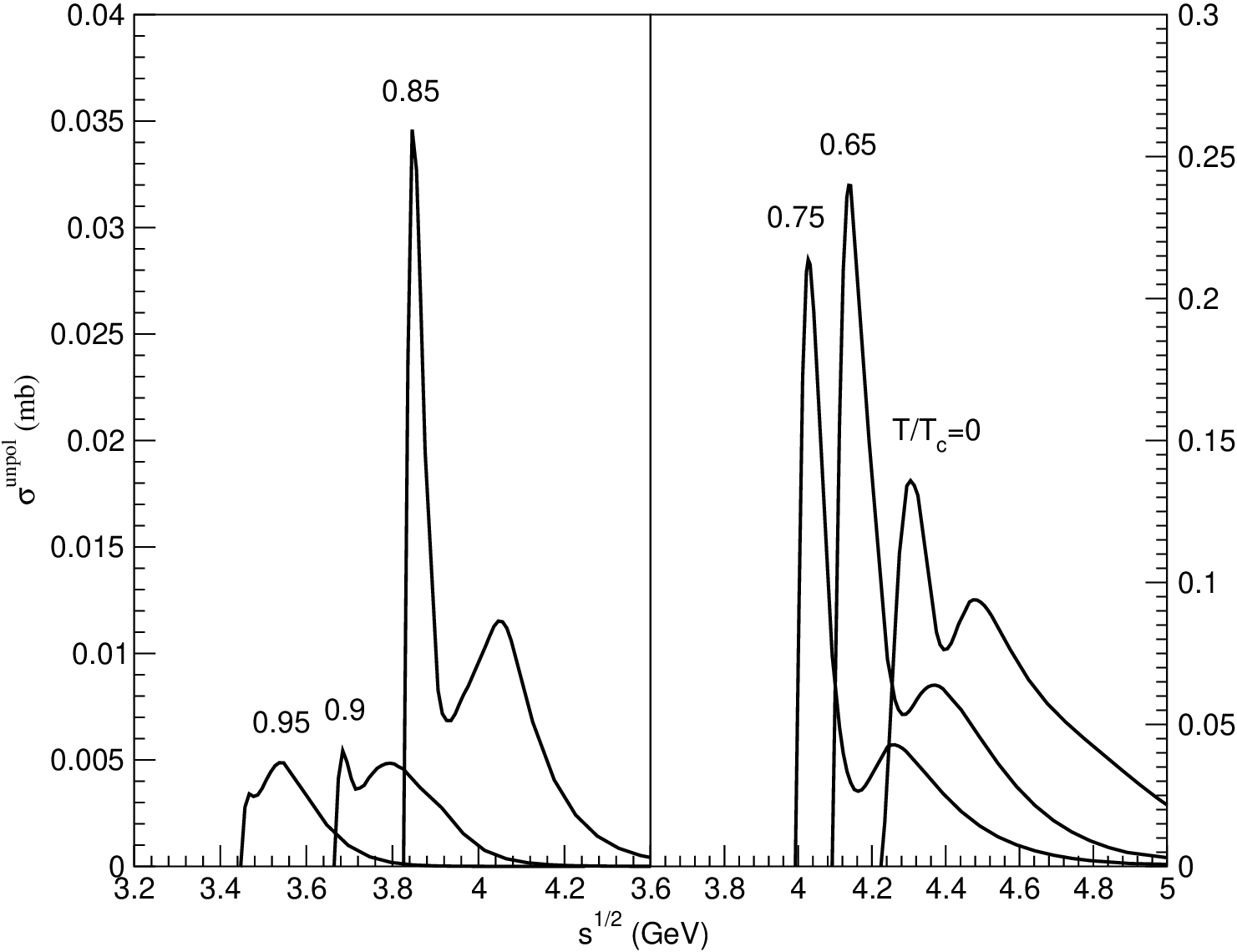}
\caption{Cross sections for $\eta+\chi_{c}\to D^{*-}_{s}+D^{*+}_{s}$ at
various temperatures.}
\label{fig23}
\end{figure}

\newpage
\begin{figure}[htbp]
\centering
\includegraphics[scale=0.65]{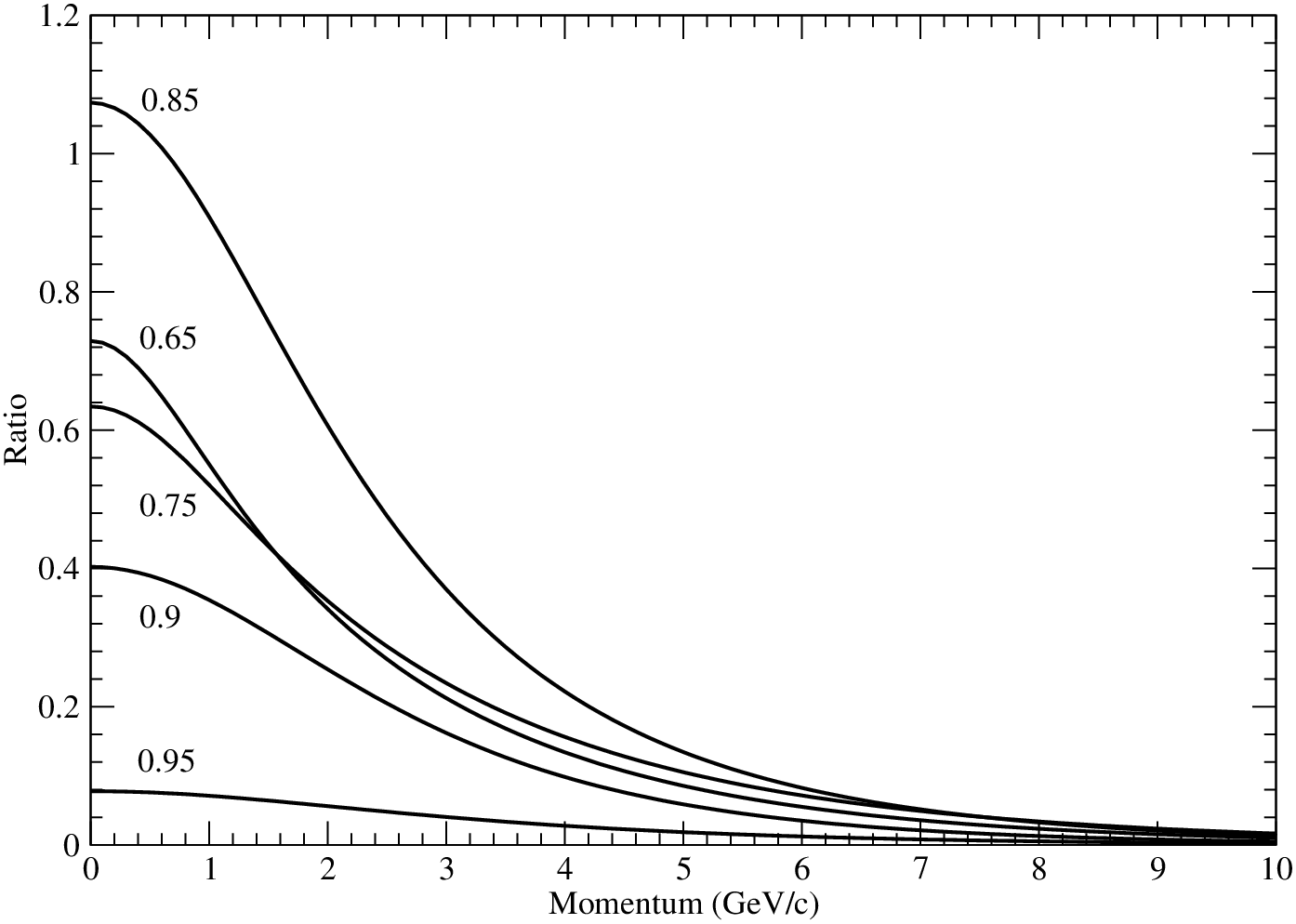}
\caption{Ratio for $\eta+J/\psi \to \bar{D}^*+D$ and 
$\pi+J/\psi \to \bar{D}^*+D$ as a function of the $J/\psi$ momentum.}
\label{fig24}
\end{figure}

\newpage
\begin{figure}[htbp]
\centering
\includegraphics[scale=0.65]{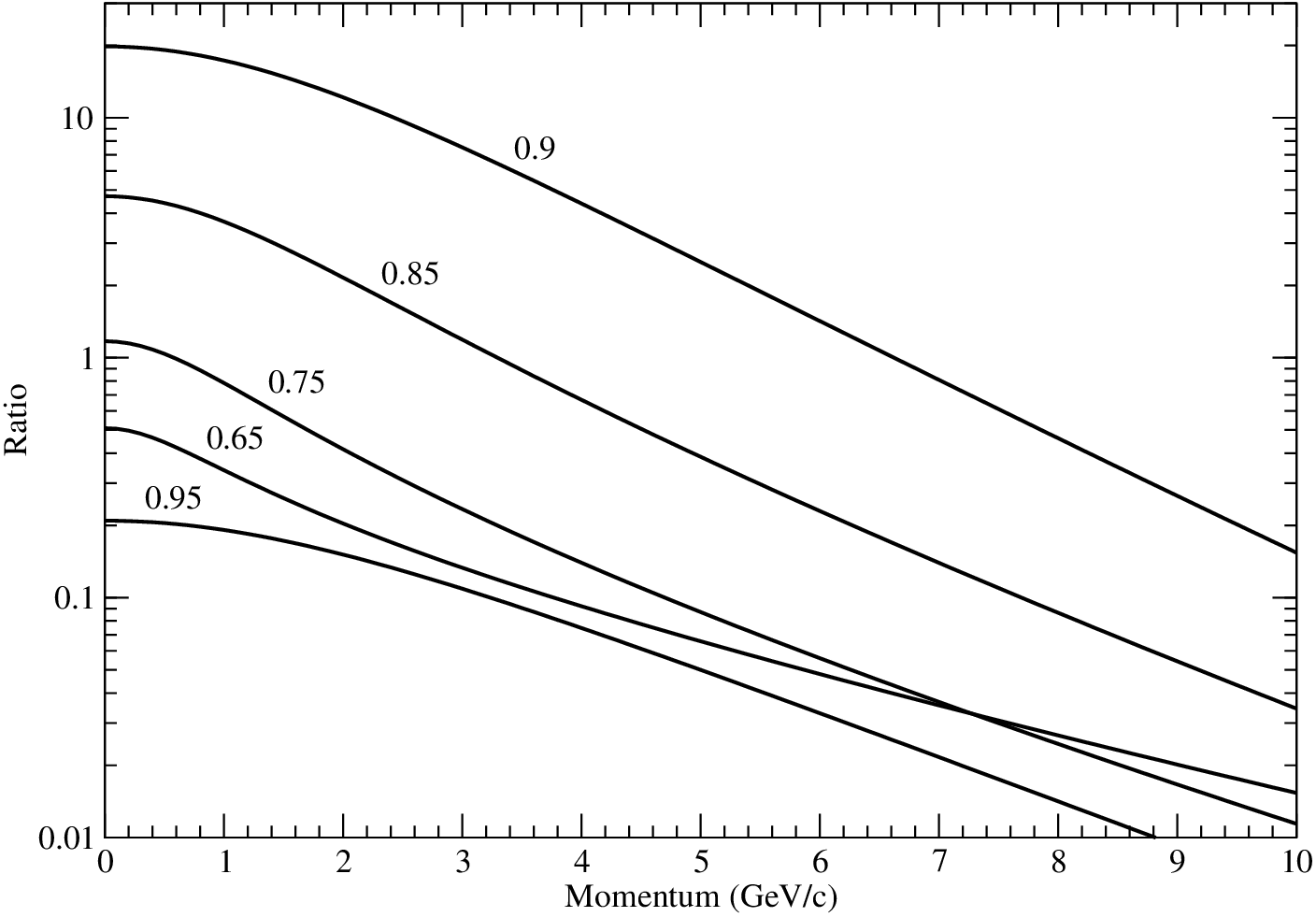}
\caption{Ratio for $\eta+J/\psi \to \bar{D}^*+D^*$ and 
$\pi+J/\psi \to \bar{D}^*+D^*$ as a function of the $J/\psi$ momentum.}
\label{fig25}
\end{figure}

\newpage
\begin{figure}[htbp]
\centering
\includegraphics[scale=0.65]{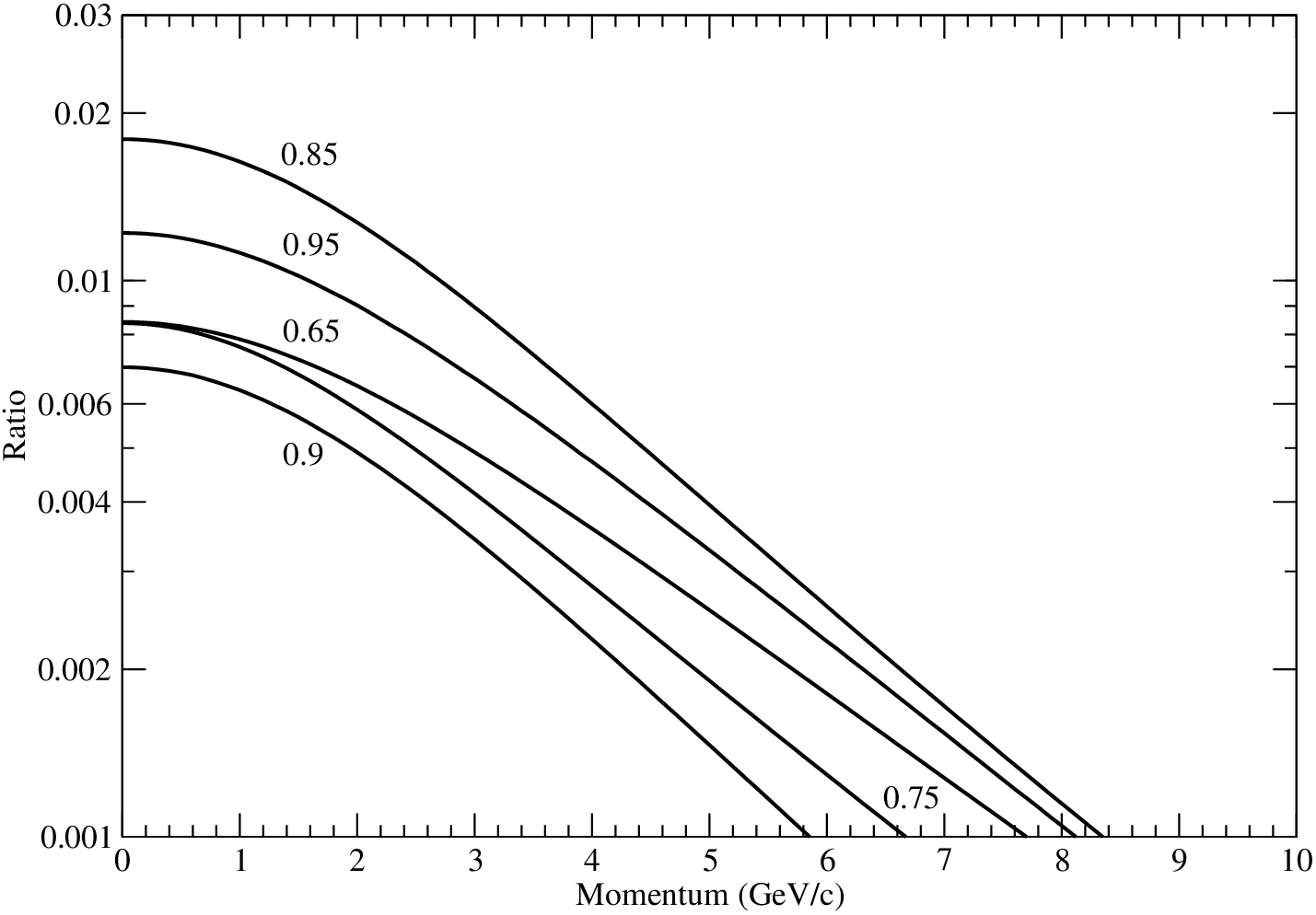}
\caption{Ratio for $\eta+\psi' \to \bar{D}^*+D$ and 
$\pi+\psi' \to \bar{D}^*+D$ as a function of the $\psi'$ momentum.}
\label{fig26}
\end{figure}

\newpage
\begin{figure}[htbp]
\centering
\includegraphics[scale=0.65]{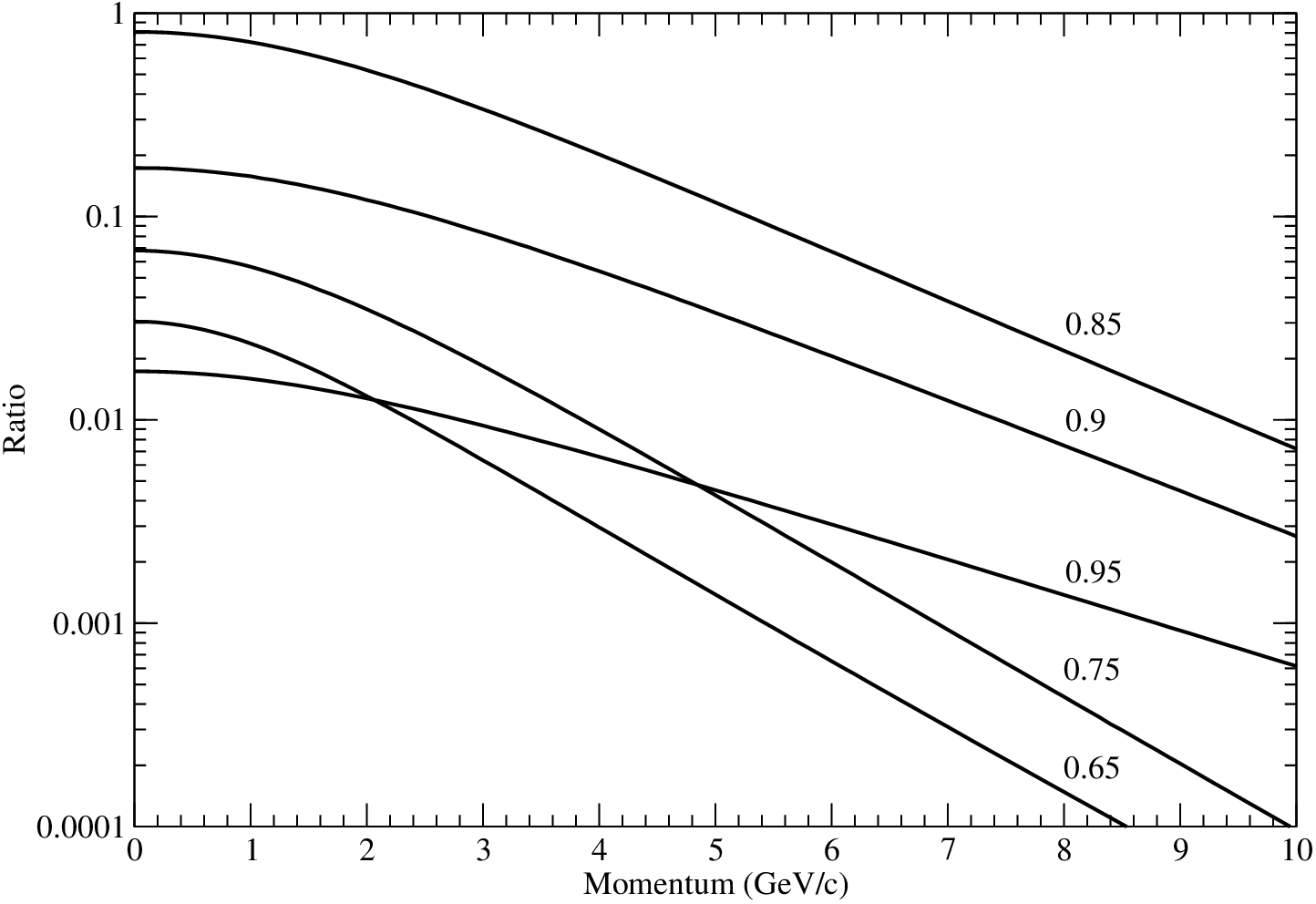}
\caption{Ratio for $\eta+\psi' \to \bar{D}^*+D^*$ and 
$\pi+\psi' \to \bar{D}^*+D^*$ as a function of the $\psi'$ momentum.}
\label{fig27}
\end{figure}

\newpage
\begin{figure}[htbp]
\centering
\includegraphics[scale=0.65]{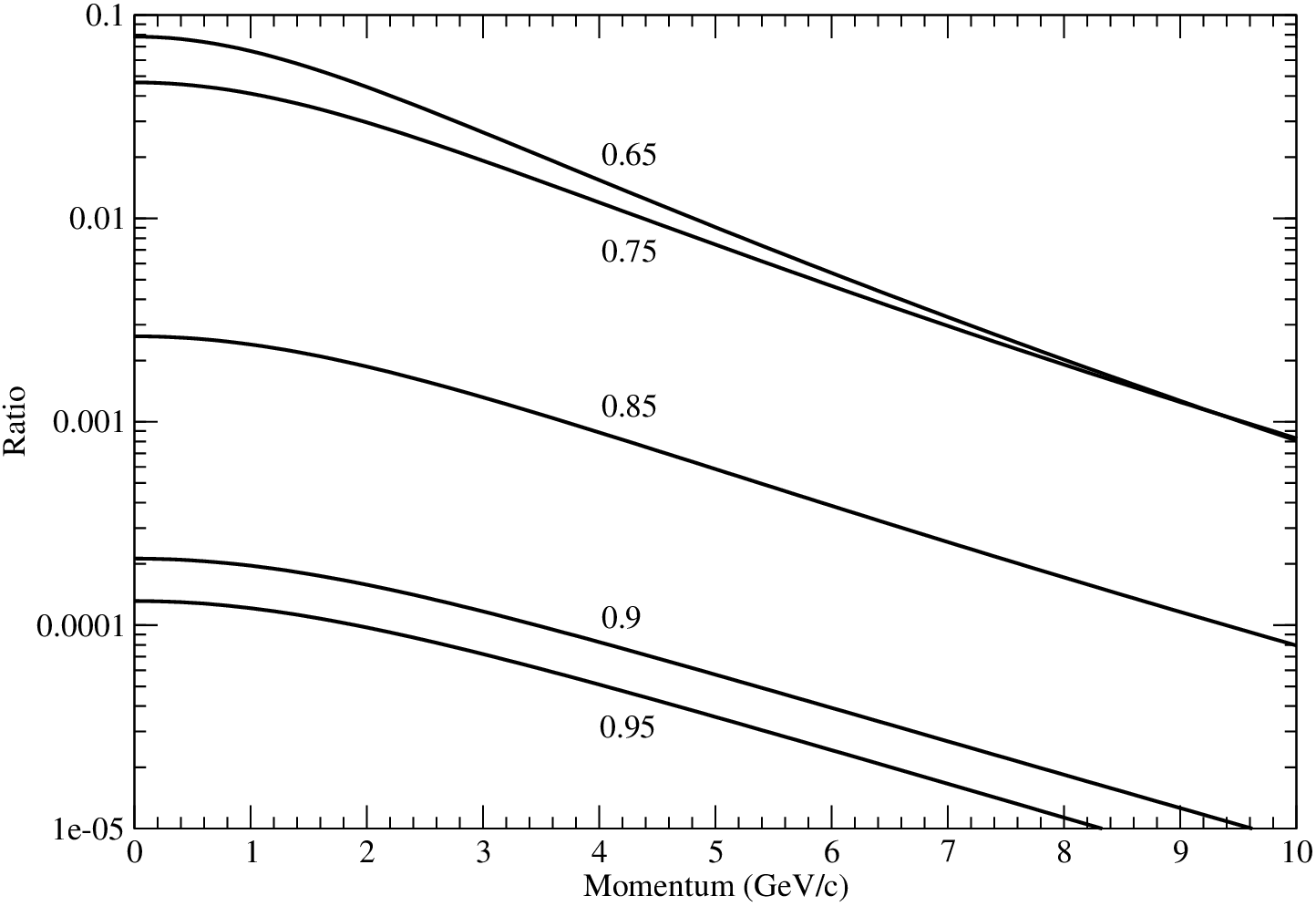}
\caption{Ratio for $\eta+\chi_c \to \bar{D}^*+D$ and 
$\pi+\chi_c \to \bar{D}^*+D$ as a function of the $\chi_c$ momentum.}
\label{fig28}
\end{figure}

\newpage
\begin{figure}[htbp]
\centering
\includegraphics[scale=0.65]{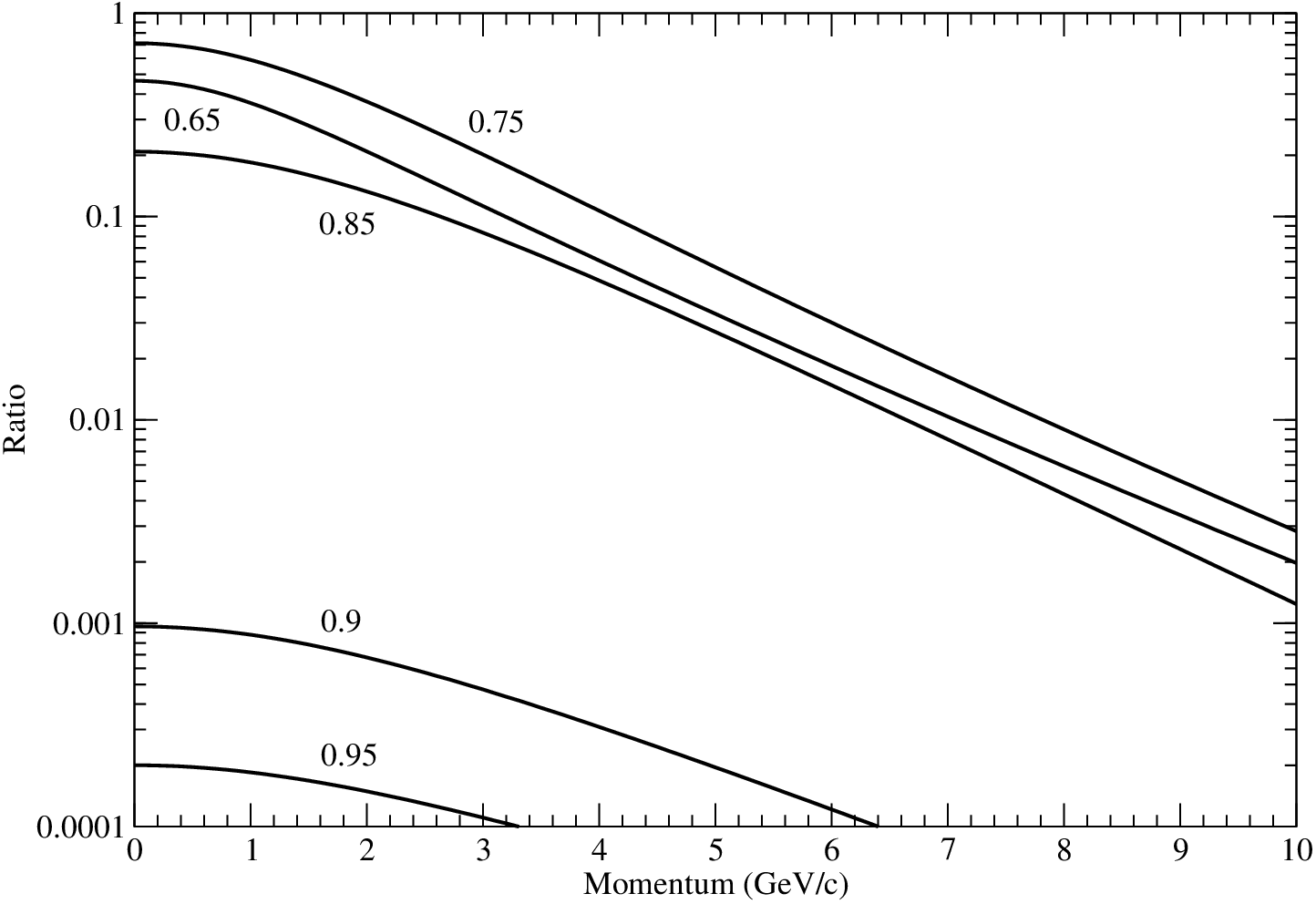}
\caption{Ratio for $\eta+\chi_c \to \bar{D}^*+D^*$ and 
$\pi+\chi_c \to \bar{D}^*+D^*$ as a function of the $\chi_c$ momentum.}
\label{fig29}
\end{figure}

\newpage
\begin{table}
\centering
\caption{Quantities relevant to the cross sections for the $KJ/\psi$ 
dissociation reactions. $a_1$ and $a_2$ are 
in units of mb; $b_1$, $b_2$, $d_0$ and $\sqrt{s_{\rm z}}$ are in units of 
GeV; $c_1$ and $c_2$ are dimensionless. }
\label{table1}
\tabcolsep=5.8pt
\begin{tabular}{llllllllll}
  \hline
  Reactions & $T/T_{\rm c}$ & $a_1$ & $b_1$ & $c_1$ & $a_2$ & $b_2$ & $c_2$ 
            &$d_0$ & $\sqrt{s_{\rm z}}$ \\
  \hline
  $KJ/\psi\to\bar{D}^*D^+_s$
  & 0     & 0.27    & 0.03  & 0.52  & 0.1     & 0.085  & 2.63 & 0.044 &4.52\\
  & 0.65  & 0.418   & 0.025 & 0.52  & 0.131   & 0.082  & 2.58 & 0.035 &4.31\\
  & 0.75  & 0.4     & 0.012 & 0.52  & 0.45    & 0.06   & 1.53 & 0.03  &4.18\\
  & 0.85  & 0.27    & 0.01  & 0.52  & 0.34    & 0.056  & 1.45 & 0.03  &4.04\\
  & 0.9   & 0.24    & 0.011 & 0.44  & 0.29    & 0.037  & 0.74 & 0.022 &3.84\\
  & 0.95  & 0.42    & 0.019 & 0.42  & 0.69    & 0.01   & 0.54 & 0.012 &3.52 \\
  \hline
  $KJ/\psi\to\bar{D}D^{*+}_s$
  & 0     & 0.22    & 0.024 & 0.55  & 0.12    & 0.081  & 2.77 & 0.04  &4.81\\
  & 0.65  & 0.23    & 0.015 & 0.55  & 0.2     & 0.066  & 2.06 & 0.034 &4.59\\
  & 0.75  & 0.23    & 0.009 & 0.51  & 0.36    & 0.048  & 1.31 & 0.03  &4.41\\
  & 0.85  & 0.235   & 0.01  & 0.51  & 0.29    & 0.049  & 1.5  & 0.029 &4.12\\
  & 0.9   & 0.135   & 0.01  & 0.48  & 0.118   & 0.044  & 1.2  & 0.02  &3.77\\
  & 0.95  & 0.25    & 0.007 & 0.59  & 0.32    & 0.011  & 0.44 & 0.0091&3.46\\
  \hline
  $KJ/\psi\to\bar{D}^{*}D^{*+}_s$
  & 0     & 0.0065  & 0.0179 & 0.53  & 0.0376 & 0.339  & 4.92 & 0.3   &5.46\\
  & 0.65  & 0.0088  & 0.075 & 0.9   & 0.03    & 0.27   & 2.76 & 0.25  &5.17\\
  & 0.75  & 0.0006  & 0.003 & 0.36  & 0.047   & 0.21   & 2.47 & 0.23  &4.94\\
  & 0.85  & 0.0092  & 0.0111& 0.5   & 0.021   & 0.2158 & 3.38 & 0.22  &4.54\\
  & 0.9   & 0.0052  & 0.016 & 0.46  & 0.0064  & 0.21   & 2.57 & 0.2   &4.31 \\
  & 0.95  & 0.215   & 0.0034& 0.52  & 0.101   & 0.0134 & 1.91 & 0.0047&3.58\\
  \hline
\end{tabular}
\end{table}

\newpage
\begin{table}
\centering
\caption{The same as Table 1 except for the $K\psi'$ dissociation.}
\label{table2}
\begin{tabular}{llllllllll}
  \hline
  Reactions & $T/T_{\rm c}$ & $a_1$ & $b_1$ & $c_1$ & $a_2$ & $b_2$ & $c_2$ 
            &$d_0$ & $\sqrt{s_{\rm z}}$ \\
  \hline
  $K\psi'\to\bar{D}^{*}D^+_s$
  & 0     & 0.038   & 0.014 & 0.45  & 0.036   & 0.214  & 2.16 & 0.019  &5.12\\
  & 0.65  & 18.3    & 0.0021& 0.42  & 13.2    & 0.013  & 1.13 & 0.0033 &3.93\\
  & 0.75  & 11      & 0.003 & 0.45  & 9.2     & 0.011  & 0.93 & 0.0059 &3.82\\
  & 0.85  & 5.52    & 0.005 & 0.55  & 2       & 0.007  & 0.39 & 0.0054 &3.68\\
  & 0.9   & 3.4     & 0.0022& 0.47  & 3.17    & 0.0089 & 1.07 & 0.0053 &3.57\\
  & 0.95  & 2.45    & 0.0019& 0.47  & 1.7     & 0.0069 & 1.07 & 0.0036 &3.4\\
  \hline
  $K\psi'\to\bar{D}D^{*+}_s$
  & 0     & 0.043   & 0.017 & 0.41  & 0.033   & 0.3    & 4.59 & 0.015  &5.09\\
  & 0.65  & 6.4     & 0.0075& 0.51  & 1.24    & 0.019  & 1.29 & 0.0091 &3.96\\
  & 0.75  & 4.5     & 0.0039& 0.53  & 2.93    & 0.016  & 1.71 & 0.0072 &3.85\\
  & 0.85  & 2.35    & 0.0036& 0.5   & 0.91    & 0.0116 & 1.5  & 0.0049 &3.78\\
  & 0.9   & 1.35    & 0.0019& 0.53  & 1.01    & 0.0083 & 1.55 & 0.0041 &3.59\\
  & 0.95  & 1.45    & 0.0014& 0.5   & 1.12    & 0.0056 & 1.5  & 0.0029 &3.36 \\
  \hline
  $K\psi'\to\bar{D}^{*}D^{*+}_s$
  & 0     & 0.495   & 0.02  & 0.63  & 0.103   & 0.011  & 0.29 & 0.023  &5.01\\
  & 0.65  & 2.3     & 0.005 & 0.51  & 1.3     & 0.018  & 1.49 & 0.0084 &4.26\\
  & 0.75  & 2.01    & 0.005 & 0.49  & 0.76    & 0.014  & 1.1  & 0.0074 &4.07\\
  & 0.85  & 0.445   & 0.0033& 0.51  & 0.208   & 0.0132 & 1.37 & 0.0055 &3.83\\
  & 0.9   & 0.214   & 0.0031& 0.41  & 0.082   & 0.0028 & 0.92 & 0.0037 &3.62\\
  & 0.95  & 1.2     & 0.0011& 0.52  & 1.11    & 0.005  & 1.55 & 0.0028 &3.34\\
  \hline
\end{tabular}
\end{table}

\newpage
\begin{table}
\centering
\caption{The same as Table 1 except for the $K\chi_c$ dissociation.}
\label{table3}
\begin{tabular}{llllllllll}
  \hline
  Reactions & $T/T_{\rm c}$ & $a_1$ & $b_1$ & $c_1$ & $a_2$ & $b_2$ & $c_2$ 
            &$d_0$ & $\sqrt{s_{\rm z}}$ \\
  \hline
  $K\chi_{c}\to\bar{D}^{*}D^+_s$
  & 0     & 0.445   & 0.104 & 2.05  & 0.162   & 0.035  & 0.55 & 0.1   &4.77\\
  & 0.65  & 2.55    & 0.049 & 1.39  & 0.83    & 0.061  & 0.71 & 0.05  &4.21\\
  & 0.75  & 4.3     & 0.029 & 1.21  & 0.8     & 0.081  & 0.62 & 0.03  &4.02\\
  & 0.85  & 3       & 0.016 & 0.72  & 0.73    & 0.021  & 2.6  & 0.018 &3.73\\
  & 0.9   & 2.1     & 0.013 & 0.65  & 0.85    & 0.017  & 1.66 & 0.014 &3.58\\
  & 0.95  & 1.86    & 0.01  & 0.78  & 0.71    & 0.021  & 1.99 & 0.014 &3.38\\
  \hline
  $K\chi_{c}\to\bar{D}D^{*+}_s$
  & 0     & 0.395   & 0.102 & 2.33  & 0.158   & 0.041  & 0.55 & 0.1   &4.98\\
  & 0.65  & 1.5     & 0.065 & 1.4   & 0.66    & 0.044  & 1.56 & 0.059 &4.19\\
  & 0.75  & 1.7     & 0.03  & 1.4   & 0.98    & 0.066  & 2    & 0.04  &4.04\\
  & 0.85  & 0.72    & 0.022 & 1.75  & 0.62    & 0.019  & 0.87 & 0.02  &3.77\\
  & 0.9   & 0.64    & 0.013 & 0.86  & 0.271   & 0.017  & 2.32 & 0.015 &3.6\\
  & 0.95  & 1.04    & 0.014 & 1.8   & 0.17    & 0.005  & 1    & 0.013 &3.36\\
  \hline
  $K\chi_{c}\to\bar{D}^{*}D^{*+}_s$
  & 0     & 0.63    & 0.09  & 3.06  & 0.09    & 0.03   & 0.68 & 0.11  &5.22\\
  & 0.65  & 0.645   & 0.042 & 1.61  & 0.152   & 0.029  & 0.51 & 0.04  &4.72\\
  & 0.75  & 0.795   & 0.03  & 1.36  & 0.188   & 0.022  & 0.45 & 0.03  &4.44 \\
  & 0.85  & 0.233   & 0.0099& 0.61  & 0.103   & 0.0238 & 2    & 0.014 &3.97\\
  & 0.9   & 0.1     & 0.0082& 0.69  & 0.045   & 0.0114 & 2.88 & 0.01  &3.73\\
  & 0.95  & 0.59    & 0.0087& 1.11  & 0.189   & 0.0139 & 4.18 & 0.01  &3.37\\
  \hline
\end{tabular}
\end{table}

\newpage
\begin{table}
\centering
\caption{The same as Table 1 except for the $\eta J/\psi$ dissociation.}
\label{table4}
\begin{tabular}{llllllllll}
  \hline
  Reactions & $T/T_{\rm c}$ & $a_1$ & $b_1$ & $c_1$ & $a_2$ & $b_2$ & $c_2$ 
            &$d_0$ & $\sqrt{s_{\rm z}}$ \\
  \hline
  $\eta$$J/\psi\to\bar{D}^{*}D$
  &  0     & 0.2    & 0.034 & 0.8   & 0.343    & 0.026  & 0.44 & 0.03 &4.18\\
  or $\bar{D}D^{*}$ &  0.65  & 0.49   & 0.021 & 0.46  & 0.317    & 0.047  & 1.2
  & 0.035 &4.04\\
  &  0.75  & 0.3    & 0.013 & 0.34  & 0.767     & 0.038  & 0.74 & 0.03 &3.94\\
  &  0.85  & 4.7    & 0.011 & 0.4   & 0.517    & 0.01   & 2.96 & 0.01 &3.69 \\
  &  0.9   & 0.407  & 0.009 & 0.6   & 0.21    & 0.011  & 0.4  & 0.01 &3.57\\
  &  0.95  & 0.18   & 0.008 & 0.5   & 0.0167   & 0.009  & 1.3  & 0.01 &3.38\\
  \hline
  $\eta$$J/\psi\to\bar{D}^{*}D^{*}$
  &  0     & 0.0213  & 0.012 & 0.53  & 0.022   & 0.27   & 9    & 0.015 &5.04\\
  &  0.65  & 0.0547  & 0.0116& 0.53  & 0.021   & 0.25   & 7.1  & 0.015 &4.68\\
  &  0.75  & 0.16   & 0.017 & 0.5   & 0.0113   & 0.03   & 3.6  & 0.02  &4.35\\
  &  0.85  & 0.83   & 0.006 & 0.48  & 0.27    & 0.017  & 0.62 & 0.01  &3.66\\
  &  0.9   & 1.57   & 0.006 & 0.5   & 0.263   & 0.011  & 1.17 & 0.005 &3.48\\
  &  0.95  & 0.407  & 0.006 & 0.54  & 0.043   & 0.02   & 1.84 & 0.005 &3.36\\
  \hline
  $\eta$$J/\psi\to D^{*-}_{s}D_{s}^+$
  & 0      & 0.11  & 0.049 & 0.7   & 0.008   & 0.02   & 0.2  & 0.05 &4.91\\
  or $D_{s}^-D^{*+}_{s}$& 0.65   & 0.147   & 0.04   & 0.7   & 0.0193   & 0.05
  & 0.23 & 0.04 &4.67 \\
  & 0.75   & 0.177 & 0.039 & 0.64  & 0.0053  & 0.001  & 0.06 & 0.04 &4.55\\
  & 0.85   & 0.13  & 0.041 & 1     & 0.0347  & 0.006  & 0.41 & 0.03 &4.37\\
  & 0.9    & 0.053  & 0.031 & 0.71  & 0.0207  & 0.02   & 0.36 & 0.03 &4.21\\
  & 0.95   & 0.033  & 0.024 & 0.47  & 0.008   & 0.02   & 0.8  & 0.02 &4.01 \\
  \hline
  $\eta$$J/\psi\to D^{*-}_{s}D^{*+}_{s}$
  & 0     & 0.0109 & 0.042 & 0.56  & 0.0293   & 0.29   & 5.5  & 0.25 &5.18\\
  & 0.65  & 0.0147  & 0.034 & 0.57  & 0.0307   & 0.27   & 3.48 & 0.24 &5.13\\
  & 0.75  & 0.0128 & 0.041 & 0.59  & 0.03    & 0.26   & 3.07 & 0.24 &5.09\\
  & 0.85  & 0.00267 & 0.03  & 0.56  & 0.0207   & 0.18   & 1.79 & 0.2  &4.82\\
  & 0.9   & 0.00087 & 0.04  & 0.55  & 0.006   & 0.18   & 2.03 & 0.19 &4.53\\
  & 0.95  & 0.003  & 0.16  & 1.72  & 0.00087  & 0.025  & 0.46 & 0.13 &4.29\\
  \hline
\end{tabular}
\end{table}

\newpage
\begin{table}
\centering
\caption{The same as Table 1 except for the $\eta \psi'$ dissociation.}
\label{table5}
\begin{tabular}{llllllllll}
  \hline
  Reactions & $T/T_{\rm c}$ & $a_1$ & $b_1$ & $c_1$ & $a_2$ & $b_2$ & $c_2$ 
            &$d_0$ & $\sqrt{s_{\rm z}}$ \\
  \hline
  $\eta\psi'\to\bar{D}^{*}D$
  &  0     & 0.019  & 0.18  & 2.9   & 0.002   & 0.013  & 0.43 & 0.2    &5.47\\
  or $\bar{D}D^{*}$&  0.65  & 0.033  & 0.084 & 1.89  & 0.0027  & 0.039  & 0.6  
  & 0.08   &4.24\\
  &  0.75  & 0.0293 & 0.06  & 1.31  & 0.0063   & 0.009  & 0.51 & 0.05   &4.1 \\
  &  0.85  & 0.0053  & 0.05  & 1    & 0.0363   & 0.01   & 0.52 & 0.01   &3.81\\
  &  0.9   & 0.0243  & 0.008 & 0.5   & 0.00133 & 0.07   & 3.1  & 0.008  &3.63\\
  &  0.95  & 0.0203  & 0.008 & 0.51  & 0.00133 & 0.02   & 1.13 & 0.0075 &3.41\\
  \hline
  $\eta\psi'\to\bar{D}^{*}D^{*}$
  &  0     & 0.0377 & 0.0115& 0.48  & 0.0153  & 0.189  & 12.5 & 0.01   &5.35\\
  &  0.65  & 2.67   & 0.0065& 0.6   & 2.33    & 0.0054 & 0.45 & 0.005  &3.98\\
  &  0.75  & 1.27    & 0.002 & 0.5   & 1.13   & 0.008  & 1.47 & 0.005  &3.81\\
  &  0.85  & 0.203  & 0.013 & 1.24  & 0.167   & 0.004  & 0.46 & 0.01   &3.67\\
  &  0.9   & 0.057  & 0.002 & 0.55  & 0.063    & 0.01   & 1.51 & 0.006  &3.53\\
  &  0.95  & 0.0257  & 0.003 & 0.47  & 0.0293  & 0.012  & 1.14 & 0.007  &3.37\\
  \hline
  $\eta\psi'\to D^{*-}_{s}D_{s}^+$
  & 0     & 0.043   & 0.011 & 0.46  & 0.0233   & 0.25   & 1.87 & 0.02   &5.13\\
  or $D_{s}^-D^{*+}_{s}$& 0.65  & 1.07   & 0.008 & 0.4   & 1.13    & 0.014
  & 0.89 & 0.01 &4.25\\
  & 0.75  & 0.6    & 0.011 & 0.92  & 0.67   & 0.008  & 0.39 & 0.01   &4.14\\
  & 0.85  & 0.233   & 0.006 & 0.48  & 0.193   & 0.011  & 0.6  & 0.01   &3.96\\
  & 0.9   & 0.068   & 0.008 & 0.63  & 0.0847  & 0.013  & 0.41 & 0.01   &3.81\\
  & 0.95  & 0.0253  & 0.003 & 0.37  & 0.0453  & 0.02   & 0.88 & 0.01   &3.68\\
  \hline
  $\eta\psi'\to D^{*-}_{s}D^{*+}_{s}$
  & 0     & 0.4     & 0.01  & 0.42  & 0.333   & 0.025  & 0.62 & 0.01   &5.03\\
  & 0.65  & 0.333   & 0.006 & 0.33  & 0.513   & 0.012  & 1.06 & 0.01   &4.48\\
  & 0.75  & 0.3     & 0.005 & 0.44  & 0.207   & 0.012  & 1.14 & 0.01   &4.33 \\
  & 0.85  & 0.057   & 0.004 & 0.45  & 0.0153  & 0.11   & 6.43 & 0.01   &4.22\\
  & 0.9   & 0.00293 & 0.005 & 0.32  & 0.0093  & 0.09   & 4.91 & 0.07   &4.28\\
  & 0.95  & 0.00407 & 0.054 & 4.43  & 0.0014  & 0.033  & 0.14 & 0.05   &4.08\\
  \hline
\end{tabular}
\end{table}

\newpage
\begin{table}
\centering
\caption{The same as Table 1 except for the $\eta \chi_c$ dissociation.}
\label{table6}
\tabcolsep=5pt
\begin{tabular}{llllllllll}
  \hline
  Reactions & $T/T_{\rm c}$ & $a_1$ & $b_1$ & $c_1$ & $a_2$ & $b_2$ & $c_2$ 
            &$d_0$ & $\sqrt{s_{\rm z}}$ \\
  \hline
  $\eta\chi_{c}\to\bar{D}^{*}D$
  & 0      & 0.107  & 0.082 & 1.1   & 0.05    & 0.033  & 0.5  & 0.05   &4.81\\
  or $\bar{D}D^{*}$& 0.65   & 0.363 & 0.03  & 0.53  & 0.0267   & 0.05   & 3.6  
  & 0.03   &4.18\\
  & 0.75   & 0.0267  & 0.069 & 1.15  & 0.193  & 0.02   & 0.52 & 0.02   &4.05\\
  & 0.85   & 0.00373 & 0.017 & 0.43  & 0.00147 & 0.005  & 0.74 & 0.01   &3.86\\
  & 0.9    & 0.000273 & 0.041 & 4.08  & 0.000004 & 0.001 & 0.57 & 0.04  &3.73\\
  & 0.95   & 0.0004 & 0.018 & 1.71  & 0.000097 & 0.003  & 0.52 & 0.02   &3.38\\
  \hline
  $\eta\chi_{c}\to\bar{D}^{*}D^{*}$
  & 0      & 0.387   & 0.096 & 2     & 0.14    & 0.035  & 0.59 & 0.1    &4.89\\
  & 0.65   & 1.57    & 0.02  & 0.7   & 1.83     & 0.04   & 1.6  & 0.03   &4.1\\
  & 0.75   & 3.13    & 0.014 & 0.66  & 1.23   & 0.006  & 0.4  & 0.0099 &3.93\\
  & 0.85   & 0.12   & 0.0026& 0.52  & 0.05    & 0.009  & 1.61 & 0.004  &3.64 \\
  & 0.9    & 0.00035& 0.006 & 0.55  & 0.000013 & 0.004 & 4.4 & 0.005   &3.55\\
  & 0.95   & 0.00053& 0.02  & 2.75  & 0.00033 & 0.01   & 0.58 & 0.02   &3.37\\
  \hline
  $\eta\chi_{c}\to D^{*-}_{s}D_{s}^+$
  & 0     & 0.54    & 0.11  & 2.62  & 0.087   & 0.038  & 0.95 & 0.11  &4.94\\
  or $D_{s}^-D^{*+}_{s}$& 0.65  & 0.807 & 0.06  & 1.55  & 0.1    & 0.1  
  & 0.56 & 0.065 &4.46\\
  & 0.75  & 0.7    & 0.05  & 1.43  & 0.06    & 0.02   & 0.53 & 0.045 &4.25\\
  & 0.85  & 0.213   & 0.02  & 0.59  & 0.0613  & 0.03   & 2.2  & 0.023 &3.94\\
  & 0.9   & 0.127   & 0.014 & 0.61  & 0.018   & 0.018  & 0.24 & 0.015 &3.81\\
  & 0.95  & 0.0693  & 0.008 & 0.55  & 0.0293  & 0.019  & 0.58 & 0.01  &3.62\\
  \hline
  $\eta\chi_{c}\to D^{*-}_{s}D^{*+}_{s}$
  & 0     & 0.08    & 0.06  & 4.68  & 0.087   & 0.2    & 0.83 & 0.08 &5.33 \\
  & 0.65  & 0.233   & 0.05  & 1.36  & 0.06    & 0.3    & 10.52& 0.045&5.04 \\
  & 0.75  & 0.193   & 0.04  & 1.4   & 0.04    & 0.28   & 11.95& 0.035&4.79\\
  & 0.85  & 0.0332  & 0.023 & 1.48  & 0.0111  & 0.24   & 2.78 & 0.02 &4.42\\
  & 0.9   & 0.0048  & 0.018 & 1.49  & 0.005   & 0.13   & 2.39 & 0.02 &4.19\\
  & 0.95  & 0.00257 & 0.013 & 1.52  & 0.00473 & 0.09   & 2.02 & 0.09 &3.87\\
  \hline
\end{tabular}
\end{table}


\begin{thebibliography}{00}
\bibitem{KS} D. Kharzeev, H. Satz, Phys. Lett. B 334, 155 (1994).
\bibitem{Peskin} M.E. Peskin, Nucl. Phys. B 156, 365 (1979); G. Bhanot, M.E. 
Peskin, Nucl. Phys. B 156, 391 (1979).
\bibitem{XKSW} X.-M. Xu, D. Kharzeev, H. Satz, X.-N. Wang, Phys. Rev. C 53, 
3051 (1996).
\bibitem{AGGA} F. Arleo, P.B. Gossiaux, T. Gousset, J. Aichelin, Phys. Rev. 
D 65, 014005 (2001).
\bibitem{MBQ} K. Martins, D. Blaschke, E. Quack, Phys. Rev. C 51, 2723 (1995).
\bibitem{BDPK} I. Bender, H.G. Dosch, H.J. Pirner, H.G. Kruse, Nucl. Phys. 
A 414, 359 (1984).
\bibitem{BS} T. Barnes, E.S. Swanson, Phys. Rev. D 46, 131 (1992); 
E.S. Swanson, Ann. Phys. (N.Y.) 220, 73 (1992).
\bibitem{WSB} C.-Y. Wong, E.S. Swanson, T. Barnes, Phys. Rev. C 62, 045201 
(2000); Phys. Rev. C 65, 014903 (2001).
\bibitem{BSWX} T. Barnes, E.S. Swanson, C.-Y. Wong, X.-M. Xu, Phys. Rev. C 68, 
014903 (2003).
\bibitem{MatinyanM} S.G. Matinyan, B. M\"{u}ller, Phys. Rev. C 58, 2994
(1998).
\bibitem{LK} Z. Lin, C.M. Ko, Phys. Rev. C 62, 034903 (2000); J. Phys. G 27,
617 (2001).
\bibitem{Haglin} K.L. Haglin, Phys. Rev. C 61, 031902 (2000);
K.L. Haglin, C. Gale, Phys. Rev. C 63, 065201 (2001).
\bibitem{OSL} Y. Oh, T. Song, S.H. Lee, Phys. Rev. C 63, 034901 (2001).
\bibitem{NNR} F.S. Navarra, M. Nielsen, M.R. Robilotta, Phys. Rev. C 64, 
021901(R) (2001).
\bibitem{MPPR} L. Maiani, F. Piccinini, A.D. Polosa, V. Riquer, Nucl. Phys. 
A 741, 273 (2004).
\bibitem{BG} A. Bourque, C. Gale, Phys. Rev. C 78, 035206 (2008); 80, 015204 
(2009).
\bibitem{Wong} C.-Y. Wong, Phys. Rev. C 65, 034902 (2002).
\bibitem{ZX} J. Zhou, X.-M. Xu, Phys. Rev. C 85, 064904 (2012).
\bibitem{PHENIX} S.S. Adler, et al., PHENIX Collaboration, Phys. Rev. C 69, 
034909 (2004).
\bibitem{ALICE} B. Abelev, et al., ALICE Collaboration, Phys. Lett. B 736, 196
(2014).
\bibitem{Sahlmuller} B. Sahlm\"{u}ller, J. Phys. G 34, S969 (2007).
\bibitem{LX} Y.-Q. Li, X.-M. Xu, Nucl. Phys. A 794, 210 (2007).
\bibitem{MottM} N.F. Mott, H.S.W. Massey, The Theory of Atomic Collisions, 
Clarendon Press, Oxford, 1965.
\bibitem{BBS} T. Barnes, N. Black, E.S. Swanson, Phys. Rev. C 63, 025204
(2001).
\bibitem{WC} C.-Y. Wong, H.W. Crater, Phys. Rev. C 63, 044907 (2001).
\bibitem{BT} W. Buchm\"{u}ller, S.-H. H. Tye, Phys. Rev. D 24, 132 (1981).
\bibitem{KLP} F. Karsch, E. Laermann, A. Peikert, Nucl. Phys. B 605, 579 
(2001).
\bibitem{ZXG} Y.-P. Zhang, X.-M. Xu, H.-J. Ge, Nucl. Phys. A 832, 112 (2010).
\bibitem{RHS} A. Rothkopf, T. Hatsuda, S. Sasaki, Phys. Rev. Lett. 108, 
162001 (2012).
\bibitem{BR} Y. Burnier, A. Rothkopf, Phys. Rev. Lett. 111, 182003 (2013).
\bibitem{LMSK} S. H. Lee, K. Morita, T. Song, C. M. Ko, Phys. Rev. D 89, 094015
(2014).
\bibitem{SX} Z.-Y. Shen, X.-M. Xu, Chin. Phys. C 39, 074103 (2015).
\bibitem{Xu} X.-M. Xu, Nucl. Phys. A 697, 825 (2002).
\bibitem{GI} S. Godfrey, N. Isgur, Phys. Rev. D 32, 189 (1985).
\bibitem{PDG} K. Nakamura, et al., Particle Data Group, J. Phys. G 37,
075021 (2010).
\bibitem{Colton} E. Colton, et al., Phys. Rev. D 3, 2028 (1971).
\bibitem{Durusoy} N. B. Durusoy, et al., Phys. Lett. B 45, 517 (1973).
\bibitem{Hoogland} W. Hoogland, et al., Nucl. Phys. B 126, 109 (1977).
\bibitem{Losty} M. J. Losty, et al., Nucl. Phys. B 69, 185 (1974).
\end{thebibliography}
\end{document}